\documentclass[a4paper, 11pt]{article}
\pdfoutput=1
\usepackage{latexsym,amsmath,amsfonts,amssymb}
\usepackage{tikz}
\usetikzlibrary{arrows,decorations.pathmorphing,cd,decorations.markings,calc}
\usepackage{mathrsfs}
\usepackage[latin1]{inputenc}
\usepackage[american]{babel}
\usepackage{graphicx}
\usepackage{bbm}
\usepackage{cite}
\usepackage[colorlinks=true, citecolor=blue, linkcolor=blue, linktocpage=true]{hyperref}
\usepackage{tcolorbox}
\usepackage{skak}
\usepackage{enumerate}
\usepackage{titlesec}

\titlespacing*{\paragraph}{0pt}{1ex plus 0.5ex minus 0.2ex}{1em}
\tikzset{snake it/.style={decorate, decoration=snake}}
\tikzset{mid arrow/.style={postaction={decorate,decoration={markings,mark=at position .5 with {\arrow{>}}}}}}

\renewcommand{\baselinestretch}{1.1}
\usepackage[a4paper,
  left=1.75cm,
  right=1.75cm,
  top=2cm,
  bottom=2cm
]{geometry}

\newcommand{\slashed}{{\bf\not}}

\newcommand{\fP}{\mathfrak{P}}

\numberwithin{equation}{section}

\newcommand{\be}{\begin{equation}} \newcommand{\ee}{\end{equation}}
\newcommand{\bea}{\begin{equation} \begin{aligned}} \newcommand{\eea}{\end{aligned} \end{equation}}

\newcommand{\cA}{\mathcal{A}}

\newcommand{\cH}{\mathcal{H}}

\newcommand{\cL}{\mathcal{L}}

\newcommand{\cM}{\mathcal{M}}

\newcommand{\cO}{\mathcal{O}}

\newcommand{\cT}{\mathcal{T}}
\newcommand{\cU}{\mathcal{U}}

\newcommand{\bC}{\mathbb{C}}

\newcommand{\bG}{\mathbf{G}}

\newcommand{\bR}{\mathbb{R}}

\newcommand{\bZ}{\mathbb{Z}}

\newcommand{\unit}{\mathbbm{1}}

\newcommand{\bx}{\mathbf{x}}

\def\repa{\raise4pt\hbox{$\square$}\mkern-14mu\raise-4pt\hbox{$\square$}}
\def\repab{\overline{\raise4pt\hbox{$\square$}\mkern-14mu\raise-4pt\hbox{$\square$}\mkern-1mu}}
\newcommand{\mon}[1]{\textbf{M} (#1)}

\DeclareMathOperator{\Tr}{Tr}

\DeclareMathOperator{\calD}{\mathscr{D}}

\makeatletter
\DeclareFontFamily{OMX}{MnSymbolE}{}
\DeclareSymbolFont{MnLargeSymbols}{OMX}{MnSymbolE}{m}{n}
\SetSymbolFont{MnLargeSymbols}{bold}{OMX}{MnSymbolE}{b}{n}
\DeclareFontShape{OMX}{MnSymbolE}{m}{n}{
    <-6>  MnSymbolE5
   <6-7>  MnSymbolE6
   <7-8>  MnSymbolE7
   <8-9>  MnSymbolE8
   <9-10> MnSymbolE9
  <10-12> MnSymbolE10
  <12->   MnSymbolE12
}{}
\DeclareFontShape{OMX}{MnSymbolE}{b}{n}{
    <-6>  MnSymbolE-Bold5
   <6-7>  MnSymbolE-Bold6
   <7-8>  MnSymbolE-Bold7
   <8-9>  MnSymbolE-Bold8
   <9-10> MnSymbolE-Bold9
  <10-12> MnSymbolE-Bold10
  <12->   MnSymbolE-Bold12
}{}

\let\llangle\@undefined
\let\rrangle\@undefined
\DeclareMathDelimiter{\llangle}{\mathopen}%
                     {MnLargeSymbols}{'164}{MnLargeSymbols}{'164}
\DeclareMathDelimiter{\rrangle}{\mathclose}%
                     {MnLargeSymbols}{'171}{MnLargeSymbols}{'171}
\makeatother

\tikzset{
  fat x/.style={
    draw=none,           
    minimum size=0.25cm,    
    inner sep=0pt,
    path picture={
      \draw[line width=2pt] 
        (path picture bounding box.south west) --
        (path picture bounding box.north east)
        (path picture bounding box.north west) --
        (path picture bounding box.south east);
    }
  }
}

\begin{document}

\thispagestyle{empty}

\vspace*{20mm}  
\begin{center}
	{\Huge   
 \textbf{When Symmetries Twist: \\[0.5em] Anomaly Inflow on Monodromy Defects}}
	\\[13mm]
   {\large Christian Copetti
        }

	\bigskip
	{\it
		  Mathematical Institute, University
of Oxford, Woodstock Road, Oxford, OX2 6GG, United Kingdom 
	}
    \bigskip
    
    {christian.copetti@maths.ox.ac.uk}
\end{center}

\bigskip

 \begin{abstract}
 \noindent
Monodromy defects describe a dynamical termination of topological symmetry operators, and are sourced by a localized background magnetic flux.
We study their properties in gapped SPT phases and, by inflow, in gapless theories with an anomalous symmetry. The background flux acts as a source for the anomaly, impacting the definition of the monodromy defect, which can only be defined as a domain wall between the symmetry generator and a topological order induced by the anomaly.
The topological dressing has several consequences, such as the presence of protected chiral edge modes on the defect's worldvolume, and the adiabatic pumping of gapless degrees of freedom bound to the localized flux.
We verify our predictions in several examples, focusing on monodromy defects for anomalous chiral symmetries, both in the continuum and on the lattice.
 \end{abstract}

\pagenumbering{arabic}
\setcounter{page}{1}
\setcounter{footnote}{0}
\renewcommand{\thefootnote}{\arabic{footnote}}

{\renewcommand{\baselinestretch}{1} \parskip=0pt
\setcounter{tocdepth}{2}
}
\newpage

\tableofcontents

\newpage 

\parindent=0pt

\section{Introduction}
The space of defects in a quantum system has been the subject of intense recent study: defects arise naturally as impurities in condensed-matter setups, and serve as probes of strongly coupled bulk dynamics.
Topological defects in particular --- i.e.\ symmetries \cite{Gaiotto:2014kfa} --- have led to a wealth of constraints on the long-distance physics, and their classification across dimensions has reached an increasing degree of maturity. Central to these developments is the notion of a 't Hooft anomaly: an obstruction to a trivial realization of the symmetry.
Much less is known about dynamical defects, which are ubiquitous in critical setups. These are often strongly coupled objects and cannot be fully captured by perturbative techniques outside of limited regimes. 

Topological and dynamical defects are, however, intimately related: symmetries naturally act on extended excitations via the appropriate ``higher" algebraic structure. Such action provides novel, non-perturbative, constraints on the physical properties of defects. 
The power of this approach has been fleshed out recently in several contexts, providing new insights into RG flows and bulk screening \cite{Antinucci:2024izg}, proposing new dualities between defect CFTs \cite{Komargodski:2025jbu}, understanding scattering amplitudes of domain walls \cite{Copetti:2024rqj,Copetti:2024dcz,Cordova:2024vsq,Cordova:2024iti} and on magnetic monopoles \cite{vanBeest:2023dbu,vanBeest:2023mbs,Loladze:2025jsq}, and constraining defect conformal data by a clever use of the Ward identities \cite{Gabai:2025zcs,Kong:2025sbk,Belton:2025ief}. 

At the same time, it has become increasingly clear that 't Hooft anomalies also lead to a rich set of consequences for the physics of dynamical defects. A prominent example of such interplay is the anomaly-induced symmetry breaking by boundary conditions \cite{Jensen:2017eof,Thorngren:2020yht}. These bulk/boundary systems present interesting anomalies in the space of boundary couplings \cite{Choi:2025ebk,Copetti:2025sym,Komargodski:2025jbu,Jones:2025khc} which lead to the pumping on nontrivial topological degrees of freedom (SPTs) as we adiabatically trace non-contractible cycles in the space of the boundary couplings.
This type of interplay between anomalies, symmetry breaking, and defects has been fleshed out in the Symmetry Breaking Exact Sequence of \cite{Debray:2023ior}, which encodes how bulk anomaly matching conditions are satisfied by defects.
Anomalies are intimately related to symmetry protected topological phases (SPTs) by the inflow mechanism, and many of these phenomena can be reinterpreted by understanding the implementation of (topological) symmetry defects in a bulk SPT.

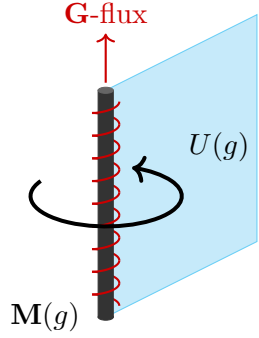
\begin{figure}[t!]
    \begin{equation*}
\begin{tikzpicture}[baseline={(0,1.5)}]
\draw[color=white!50!cyan, fill=white!80!cyan] (0,0) -- (2,1) -- (2,4) -- (0,3) -- cycle;
\node at (1.5,2.25) {$U(g)$};
  \foreach \i in {0,...,8} {
      \draw[red!80!black, thick, domain=0:180, samples=30] plot ({0.18*cos(\x)}, {0.3*\i + 0.15 + 0.3*(\x/360) + 0.05*sin(\x)});
  }
  \filldraw[black!80] (-0.1,0) rectangle (0.1,3);
  \filldraw[black!80] (0,0) ellipse (0.1 and 0.04);
  \filldraw[black!60] (0,3) ellipse (0.1 and 0.04);
  
  \foreach \i in {0,...,8} {
      \draw[red!80!black, thick, domain=180:360, samples=30] plot ({0.18*cos(\x)}, {0.3*\i + 0.15 + 0.3*(\x/360) + 0.05*sin(\x)});
  }
  \draw[red!80!black, thick, ->] (0,3.1) -- (0,3.75) node[above] {$\bG$-flux};
\node at (-0.8,0) {$\mon{g}$};
\draw[line width=1.5,->] (-0.86, 1.8) arc (150:430:1.0 and 0.4); 
\end{tikzpicture}
\end{equation*}
\caption{A monodromy defect $\mon{g}$ can be thought of as a dynamical termination of a symmetry operator $U(g)$, which imposes twisted boundary conditions in the angular direction. Alternatively, it is sourced by a localized $\bG$-flux (fractional, in the discrete case).}
\label{fig: monodromydef}
\end{figure}

In this paper, we explore yet another scenario. 
In a symmetric system, the spectrum of dynamical defects is vastly enriched. Besides the symmetry defects themselves, the natural objects to consider are the so-called \emph{Twist} or \emph{Monodromy} defects, denoted by $\mon{g}$, $g \in \bG$.
Pictorially, they describe the consistent ways to terminate a bulk symmetry defect $U(g)$ (see Figure \ref{fig: monodromydef}).
Physically, the monodromy defect can be induced by background $\bG$ flux localized on the $\mon{g}$ worldvolume. This allows to externally dial the group parameter $g$ and explore the space of monodromy defects adiabatically.
Monodromy defects are generically non-topological\footnote{A topological monodromy defect implies that the bulk symmetry does not act faithfully, and is implemented by a condensation defect \cite{Roumpedakis:2022aik,Copetti:2023mcq}. This type of defects has been studied from several angles in the context of topological orders \cite{Barkeshli:2014cna,Antinucci:2022vyk}.} and have been studied in several settings, including supersymmetric theories \cite{Gukov:2008sn,Cordova:2017mhb,Bianchi:2019sxz,Arav:2024exg,Arav:2024wyg}, free field theories \cite{Bianchi:2021snj}, vector models \cite{Yamaguchi:2016pbj,Soderberg:2017oaa,Giombi:2021uae,SoderbergRousu:2023pbe,Gimenez-Grau:2022czc,Kravchuk:2025evf}, the 3d Ising model \cite{Billo:2013jda,Gaiotto:2013nva} and condensed matter/lattice systems \cite{Barkeshli:2014cna,wang2021scaling,wang2022scaling,Fidkowski:2016svr,Barkeshli:2025cjs}. 
We will be concerned with understanding the physics of the monodromy defects $\mon{g}$ for an anomalous bulk $\bG$ symmetry.\footnote{In most of the paper $\bG$ is an ordinary (0-form) group. We will briefly explain the generalization to higher-form symmetries where needed.}
The first step will be to study the physics of (topological) monodromy defects $\mathbf{T}(g)$ in a bulk symmetry protected topological (SPT) phase:
\be
\Omega \in \mathbf{SPT}^{d+1}_{\bG} \, .
\ee
This is a very natural setup: different types of defects provide natural probes into bulk phases of matter. For example, the so-called \emph{pinning-field} defects \cite{Assaad:2013xua,Popov:2025cha} can probe symmetry-breaking phases in a local manner. 
SPT phases are most often probed by the study of their protected boundary modes \cite{Hasan:2010xy,Chen:2011pg}, but, especially in (1+1) dimensions, a similar understanding can be obtained by considering the so-called string order parameters \cite{denNijs:1989,Kennedy:1992,Pollmann:2009mhk}. Their higher-dimensional generalizations are exactly the topological monodromy defects $\mathbf{T}(g)$. 
We will show that, in a bulk SPT, $\mathbf{T}(g)$ will generically carry a set of protected edge modes.
More precisely, for a $p$-form symmetry we find the maps:
\be
 \begin{tikzcd}[row sep=1.5em, column sep=huge]
 \mathbf{SPT}_{\bG}^{d+1} \arrow[r, "\tau"] & \mathbf{SPT}_{C_{\bG}(g)}^{d-p} \arrow[r, "\partial"] & \mathbf{TFT}^{d-p-1}_{C_{\bG}(g)} \\
 \Omega \arrow[r, |->] & \tau(g) \arrow[r, |->] & \mathbf{T}(g)
 \end{tikzcd}
 \ee
where $\mathbf{TFT}^{d-p-1}_{C_{\bG}(g)}$ is the space of $C_{\bG}(g)$-symmetric Topological Field Theories and $C_\bG(g)$ is the centralizer of $g$ --- the symmetry preserved by the monodromy defect.\footnote{The slant by a $p$-form parameter shifts the cohomological degree by $p+1$, which gives the dimensions in the diagram. For $p=0$ this reduces to the textbook statement $H^{d+1}(B\bG) \to H^d(BC_\bG(g))$.} For most of the paper we will be interested in the case $p=0$.
Physically, the first map amounts to compactification on a $g$-twisted circle (or $T^{p+1}$, for $p$-form $\bG$) and is known as the \emph{transgression} or \emph{slant product}, while the second describes a topological boundary of the SPT $\tau(g)$.
We will often leave the superscripts $d$ and $p$ implicit to lighten the notation. 

\begin{figure}
    \centering
    \scalebox{0.9}{
   \begin{tikzpicture}[baseline={(0,1.25)}, x={(1.2cm,-0.3cm)}, y={(0.4cm,0.5cm)}, z={(0cm,1.2cm)}]
        \filldraw[gray!10, opacity=0.7, draw=gray!80!black, thick, line join=round] plot[domain=0:180, samples=50] ({2*cos(\x)}, {2*sin(\x)}, 0) -- plot[domain=180:0, samples=50] ({2*cos(\x)}, {2*sin(\x)}, 2) -- cycle;
        \draw[gray!80!black, thick, opacity=0.7] ({2*cos(26.57)}, {2*sin(26.57)}, 0) -- ({2*cos(26.57)}, {2*sin(26.57)}, 2);
        
        \foreach \ang in {60, 90, 120, 150} {
            \draw[gray!50!black, thin, opacity=0.3] ({2*cos(\ang)}, {2*sin(\ang)}, 0) -- ({2*cos(\ang)}, {2*sin(\ang)}, 2);
        }
        \filldraw[gray!10, opacity=0.8, draw=gray!80!black, thick, line join=round] (0,0,2) -- plot[domain=0:180, samples=50] ({2*cos(\x)}, {2*sin(\x)}, 2) -- cycle;

        \filldraw[blue, opacity=0.5, draw=blue!80!black, thick, line join=round] (0,0,0) -- (-2,0,0) -- (-2,0,2) -- (0,0,2) -- cycle;
        \filldraw[blue!10, opacity=0.7, draw=blue!50!black, thick, line join=round] (0,0,0) -- (2,0,0) -- (2,0,2) -- (0,0,2) -- cycle;

        \draw[ultra thick, black] (0,0,0) -- (0,0,2);
        
        \node[black] at (1,0,1) {$U(g)$};
        \node[black] at (-1,0,1) {$\mathbf{T}(g)$};
        \node[below] at (0,0,0) {$\mon{g}$};
        \node[black] at (0,1,2) {$\tau(g)$};
    \end{tikzpicture}
    } \qquad \qquad \qquad
    \scalebox{0.9}{
    \begin{tikzpicture}[scale=1.5, baseline={(0,0)}]
        \draw[thick, fill=gray!10] (0,0) circle (1);
        
        \draw[dashed, gray] (1,0) arc (0:180:1 and 0.2);
        \draw[gray] (-1,0) arc (180:360:1 and 0.2);

        \draw[blue, ultra thick] (0,1) arc (90:270:0.4 and 1);
        \draw[cyan, ultra thick] (0,1) arc (90:-90:0.4 and 1);

        \filldraw[black] (0,1) circle (0.05) node[above] {$\mathbf{M}(g)$};
        \filldraw[black] (0,-1) circle (0.05) node[below] {$\mathbf{M}(g)$};

        \node[black] at (-0.7, 0) {$\mathbf{T}(g)$};
        \node[black] at (0.7, 0) {$U(g)$};
     \node at (1.375,0) {$\times$};
        \draw[->, thick, black] (1.75, -0.8) -- (1.75, 0.8) node[above] {$\mathbb{R}$};
    \end{tikzpicture}
    }
    \caption{Left: The monodromy defect $\mathbf{M}(g)$ defined as a domain wall between the symmetry defect $U(g)$ and the topological decoration $\mathbf{T}(g)$. Right: The radial quantization frame $S^2 \times \mathbb{R}$ around a point on the monodromy defect. The surfaces $U(g)$ and $\mathbf{T}(g)$ intersect the spatial $S^2$ slice along meridians meeting at the poles.}
    \label{fig: radialquant}
\end{figure}
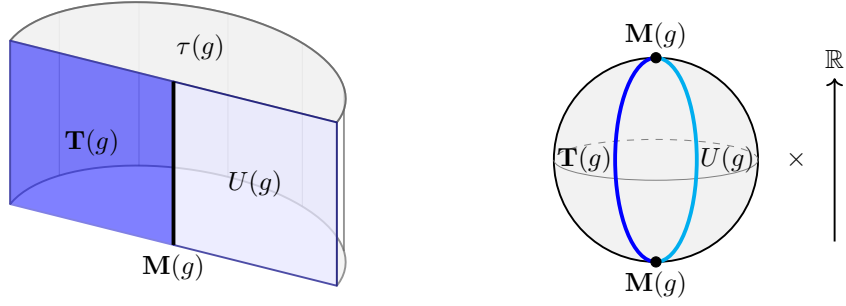

The second part of the paper instead deals with anomalous (gapless) bulk theories and their monodromy defects $\mon{g}$. We leverage our previous results and describe them as edge modes of our SPT setup.
A symmetry defect for an anomalous symmetry generally describes an interface between the starting theory, and the theory stacked with an SPT $\tau(g)$. 
The naive definition --- terminating $U(g)$ on $\mon{g}$ --- is inconsistent, since the two sides of $U(g)$ are no longer the same theory.
Our central observation is that the obstruction is resolved by introducing a second interface $\mathbf{T}(g)$ between the SPT $\tau(g)$ and the vacuum, and defining the monodromy defect as the \emph{domain wall} between $U(g)$ and $\mathbf{T}(g)$, see Figure \ref{fig: radialquant}.
The interface $\mathbf{T}(g)$ is in general non-invertible, and its topological data control the protected sectors localized on $\mon{g}$.
 In particular, if a topological domain wall does not exist, the monodromy defect as such is ill-defined.
Using these methods, we are led to a natural definition of the twisted Hilbert space $\cH_g$ in radial quantization, see Figure \ref{fig: radialquant}.
Importantly, $\mathbf{T}(g)$ cannot be placed on a manifold with boundary without leaving behind protected edge modes on $\mon{g}$.
In $(3+1)d$, $\mathbf{T}(g)$ can be a $(2+1)d$ chiral topological order, whose anomalous boundary enforces a chiral $(1+1)d$ sector on $\mon{g}$.
Such modes are protected by the bulk 't Hooft anomaly and are thus resilient. 
A related phenomenon has recently been observed for the monodromy of a non-invertible duality symmetry in $(3+1)d$ Maxwell theory \cite{Shao:2025qvf,Bashmakov:2026yuo}.
For free $(3+1)d$ Dirac fermions we make the present statement precise: the axial monodromy defect supports a chiral $(1+1)d$ sector, pumped in by adiabatic flux insertion.
We expect analogous chiral edge modes on the monodromy defects of $(3+1)d$ QED.

In this presentation, the action of a bulk symmetry $U(h) \in C_{\bG}(g)$ on the monodromy defect $\mon{g}$ is encoded in nontrivial topological junctions localized on $\mathbf{T}(g)$.
This allows to directly link the symmetry action on $\mon{g}$ to a computation in the TFT $\mathbf{T}(g)$.
We also study the imprint of the bulk 't Hooft anomaly on the global structure of the space of monodromy defects $\{\mon{g}\}_{g \in \bG}$. This is captured by an anomaly in the space of defect couplings \cite{Cordova:2019jnf,Cordova:2019uob,Choi:2025ebk,Komargodski:2025jbu,Copetti:2025sym}, which obstructs a globally consistent trivialization of the family. On $\mon{g}$ this implies that pumping one unit of magnetic flux through the thin solenoid binds it to a set of protected edge modes. This realizes the Callan--Harvey anomaly inflow mechanism \cite{Callan:1984sa} in a setting where the bulk is itself gapless: the modes localized on $\mon{g}$ saturate the bulk anomaly in essentially the same way as in the gapped case.

\paragraph{Structure of the paper} Let us outline the structure of the paper. In Section \ref{sec: monsym} we describe the general properties of monodromy defects and the definition of the twisted Hilbert space in a continuum QFT.
In Section \ref{sec: topomon} we consider (topological) monodromy defects $\mathbf{T}(g)$ as probes of bulk SPTs and describe their salient properties: protected edge modes and nontrivial SPT pumping.
In Section \ref{sec: gapless} we extend our results to gapless anomalous systems and their monodromy defect $\mon{g}$ by studying the edge modes of the SPT setup. These lead to the prediction of protected (and often gapless) sectors on $\mon{g}$.
In Section \ref{sec: chiralmon} we study the theory of free (3+1)d Dirac fermions. We point out how adiabatically pumping a quantum of flux leaves behind a (decoupled) (1+1)d chiral sector on $\mon{g}$ and connect this prediction with the multi-valuedness of the defect $b$-function \cite{Bianchi:2021snj} and compare this with the case of a \emph{vector-like} monodromy defect (i.e.\ associated to the vector $U(1)$).  
In Section \ref{sec: lattice} we describe monodromy defects in a lattice setup, and test again our predictions with different methods. We conclude with a short outlook of future prospects. We provide more applications of our results in the Appendix. In Appendix \ref{sec: 1p1} we study how our results are realized in the (1+1)d compact boson, and in Appendix \ref{sec: 2Group} we showcase an application to multiflavor bosonic QED in (2+1)d.

 \paragraph{Author's Note:} While this work was being finalized were informed that related ideas are being pursued in \cite{gabriel}.

\section{Monodromy defects and their symmetries} \label{sec: monsym}
We start by reviewing the definition of monodromy defects $\mon{g}$ and how global symmetries are realized in their presence. This will serve to set notation and ideas for the remainder of the paper. Parallel discussions can be found in several recent works \cite{Giombi:2021uae,Bianchi:2021snj,Kravchuk:2025evf}. A discussion in the context of quantum lattice models is deferred to Section \ref{sec: lattice}.
We will focus mainly on the case of a 0-form symmetry $\bG$, which can be either discrete or continuous.

\subsection{Monodromy defects}
As explained in the introduction, monodromy $\mon{g}$ defects can be understood as implementing (generically non-topological) terminations of bulk symmetry defects $U(g)$.
Alternatively, they describe a system on which twisted boundary conditions around the symmetry surface are imposed. Employing cylindrical coordinates around $\mon{g}$
\be \label{eq: cylindricoord}
ds^2 = d \rho^2 + \rho^2 d \theta^2 + h_{ab} dy^a dy^b \, ,
\ee
where $h_{ab}$ is the induced metric on the defect's worldvolume $\Sigma$, a fundamental field $\Phi$ in a representation $R_g$ of $\bG$ satisfies:
\be
\Phi(\rho, \theta + 2 \pi, y^a) = R_g^{-1} \cdot \Phi(\rho, \theta, y^a) \, .
\ee
The appearance of $R_g^{-1}$ stems from our conventions for the orientation of the defect.
For continuous $\bG = U(1)$ we write $g = e^{2 \pi i \nu Q}$, with $Q$ the $U(1)$ charge generator and $\nu \in [0,1)$ the \emph{twist parameter}; $\nu$ will play a central role throughout the paper.
In a critical setup, the $S^1$ spanned by $\theta$ can be made of fixed radius by a conformal transformation. In these coordinates, the space of states underlying the twisted field configurations differs from the Hilbert space of the theory without defect, and is often denoted by $\cH_g$ (the twisted Hilbert space).

We can also describe the monodromy defect as a disorder operator by introducing a locally flat, singular background gauge field $A$ for $\bG$ with a prescribed monodromy around any circle $C$ linking $\mon{g}$:
\be
\exp\left( i \oint_C A \right) = g \in \bG \, .
\ee
Notice that, for a non-abelian $\bG$, this breaks the symmetry group to the centralizer $C_{\bG}(g)$ of $g$.\footnote{Recall that the centralizer $C_{\bG}(g)$ of $g$ is defined as the set of elements $h \in \bG$ such that $h^{-1} g h = g$.}
From this perspective, the monodromy defect constitutes an infinitely thin solenoid which is threaded by magnetic background flux for the bulk $\bG$ gauge field. Dialing the strength of the magnetic field describes different ways to ``twist" the system and allows to adiabatically interpolate between monodromy defects for different $U(g)$s. Such interpolation allows us to describe a ``family" of monodromy defects, labeled by the parameter $g \in \bG$. Notice however that, contrary to most ``family-style" studies, the tuning of $g$ does not correspond to deformation by a \emph{local} operator on $\mon{g}$.\footnote{However, monodromy defects are still bona-fide local dynamical objects, as they have e.g. a well defined displacement operator and defect OPE.}

 The definition of a monodromy defect is far from unique. Performing defect fusion with any genuine defect leads to new monodromy defects. In practice, we will implicitly assume that the defects studied are in some sense ``minimal", for example they minimize the appropriate defect central charge.

The concept of monodromy defects can be generalized to higher-form symmetries $\bG^{(p)}$, describing a $(d-p-2)$-dimensional dynamical defect where the $(p+1)$-form background gauge field $B^{(p+1)}$ has prescribed monodromy on the linking sphere $S^{p+1}$ around $\mon{g}$:
\be
\exp\left( i \oint_{S^{p+1}} B^{(p+1)} \right) = g_p \in \bG^{(p)} \, .
\ee
 Such defects do not describe twisted boundary conditions for \emph{local} operators, but rather for extended, charged probes, such as Wilson lines. In this paper we will mainly focus on 0-form symmetries, but many of our findings generalize easily to higher-form symmetries.

\paragraph{Conformal map and the defect Hilbert space} For a critical theory, there are several different Weyl frames which make the operator content of the twist defect manifest. A particularly natural one is the ``radial quantization" frame, which leverages the state-operator correspondence of a critical bulk system.

In practice, one performs a conformal mapping to $S^{d-1} \times \bR$. The $\bR$ factor represents time, and the remaining defect directions occupy an $S^{d-3}$ inside the $S^{d-1}$. Precisely we have:
\be
\rho = e^\tau \sin(u) \, , \ \ \ y^a = e^{\tau} \cos(u) n^a \, ,
\ee
where $n^a$ is a unit vector on $S^{d-3}$. The metric is (removing a Weyl factor of $e^{2 \tau}$):
\be
ds^2 =  d \tau^2 + du^2 + \cos^2(u) d \Omega_{d-3}^2 + \sin^2(u) d \theta^2  \, . 
\ee
Infinite past corresponds to the point $\rho = y^a=0$ on the defect worldvolume, so by the state-operator correspondence the partition function on this geometry computes the spectrum of local defect operators. For $d=(1+1)$, this recovers the radial quantization on the cylinder, while for $d=(3+1)$ the $S^3$ is parametrized in Hopf coordinates:
\be
ds^2 = d \tau^2 + du^2 + \cos^2(u) d \phi^2 + \sin^2(u) d \theta^2  \, .
\ee
Notice that fields can be mildly singular (but still square-integrable) at the defect's location $u \sim 0$. This conformal map allows us to define a Hilbert space of states on $\mon{g}$, the \emph{twisted} Hilbert space $\cH_g$.

\subsection{Global symmetries of $\mon{g}$}
We now discuss the symmetry preserved by a monodromy defect $\mon{g}$.

\paragraph{Internal Symmetry.} If $\bG$ is a continuous non-abelian group its current $J^\mu$ is itself twisted by the boundary conditions. Thus, the only surviving bulk currents in the presence of $\mon{g}$ are those which commute with (the adjoint action of) $g$.
This is the avatar of a well-known phenomenon: in the presence of monodromy the symmetry group preserved by the defect is reduced to the centralizer
\be
C_{\bG}(g) \subseteq \bG \, .
\ee
If the symmetry is discrete, the same conclusion can be reached by considering the action of a bulk symmetry operator $U(h)$ on the defect system, which sends $U(g)$ to $U(h^{-1} g h)$.
Clearly, if $h$ does not lie in the centralizer, the defect $\mon{g}$ is mapped into $\mon{h^{-1} g h}$.\footnote{In other words, the family ${\mon{h^{-1}gh}}_{h \in \bG}$ is permuted by the bulk $\bG$-action. In this work, however, we largely focus on the symmetry of a given fixed defect.}

Notice that these observations do not say anything about whether $C_{\bG}(g)$ is broken by the monodromy defect itself. The precise breaking pattern depends on the microscopic definition of $\mon{g}$: for example, we could trigger symmetry breaking by turning on a relevant defect operator on $\mon{g}$. 
Furthermore, the breaking $\bG \to C_\bG(g)$ is not localized on $\mon{g}$, but sourced by the insertion of the topological defect $U(g)$. As such, it does not give rise to a well-defined tilt operator \cite{Padayasi:2021sik,Cuomo:2021cnb,Drukker:2022pxk,Herzog:2023dop} on the $\mon{g}$ worldvolume:
\be
d \star J \neq t(y^a) \delta(\Sigma) \, .
\ee
Consequently monodromy defects for different $g$ cannot be reached from one another by a localized defect deformation. In this sense the family ${\mon{g}}_{g\in\bG}$ is rather different from the standard families of symmetry-breaking defects studied in the literature.

\paragraph{Spacetime symmetry.} Just as interesting is the interplay between the monodromy and the spacetime symmetry in the presence of the defect. Recall that a codimension-2 conformal defect preserves a subgroup:
\be
SO(d-2,2) \times SO(2) \, ,
\ee
of the bulk conformal group. The first factor describes the conformal symmetry on the defect's worldvolume, while the second encodes transverse rotations. In the case of a monodromy defect $\mon{g}$, while the first factor is obviously preserved, the second is trickier to discuss. 
A naive picture is that a transverse rotation also moves the attached topological surface $U(g)$, which can then be undone by its topological invariance. This is only correct up to the $\bG$-action on charged fields, as we now make precise.

For a fixed $g \in \bG$ we can always go into an appropriate Cartan basis and represent the twist as a $U(1)$ rotation by an angle $2 \pi \nu$. In this case the (irreducible) components of $\Phi$ transform under a $2\pi$ transverse rotation as:
\be
\Phi(\rho, \theta + 2 \pi, y^a) = e^{2 \pi i (-\nu n_\Phi + s)} \Phi(\rho, \theta, y^a) \, ,
\ee
where $s$ is the spin of $\Phi$ under the transverse $SO(2)$ and $n_\Phi$ is its $\bG$-charge.
This immediately implies that the (naive) SO(2) quantum number $s$ of $\Phi$ is quantized as:
\be \label{eq: spinquant}
s \in -n_\Phi \nu + \bZ \, .
\ee
Equivalently, the transverse $SO(2)$ rotation gets entangled with the bulk $\bG$ symmetry: a $2\pi$ rotation acts on charged fields as $e^{-2\pi i \nu Q}$.
For discrete $\bG = \bZ_n$ this realizes the physical rotation as an $n$-fold cover of $SO(2)$,
\be
1 \longrightarrow \bZ_n \longrightarrow SO(2)_{\rm phys} \longrightarrow SO(2) \longrightarrow 1 \, ,
\ee
reflecting the fractional quantization of $s$ in eq.\ \eqref{eq: spinquant}.
For continuous $\bG$, the mixing is unobstructed and can be undone by passing to the redefined generator
\be
\widehat{Q}_{SO(2)} = Q_{SO(2)} + \nu Q \, ,
\ee
which generates a compact $SO(2)$.

\section{Monodromy defects as SPT probes} \label{sec: topomon}
In this Section, we consider the definition of topological monodromy defects in a bulk Symmetry Protected Topological (SPT) phase. We will denote these objects by $\mathbf{T}(g)$, to distinguish them from their dynamical counterparts $\mon{g}$ to which they will be related by inflow. We will connect this discussion with the one of anomalous gapless theories in Section \ref{sec: gapless} by considering edge modes of the SPT analysis.

However, there is also a physical motivation to consider this setup by itself: the monodromy defects $\mathbf{T}(g)$ allow to \emph{probe} the bulk SPT without resorting to boundary edge modes, by considering the response of the system to localized (background) fluxes instead. These properties provide natural generalizations of ``string'' order parameters considered in the SPT literature.

\subsection{Decorated symmetry defects and slant product}
Let us start by considering a bulk SPT
\be
\Omega \, \in \, \mathbf{SPT}^{d+1}_{\bG} \, ,
\ee
This is an almost trivial theory and its Hilbert space is one-dimensional on every closed manifold $Y^{d+1}$. However, the realization of the $\bG$ symmetry in this theory is quite rich.

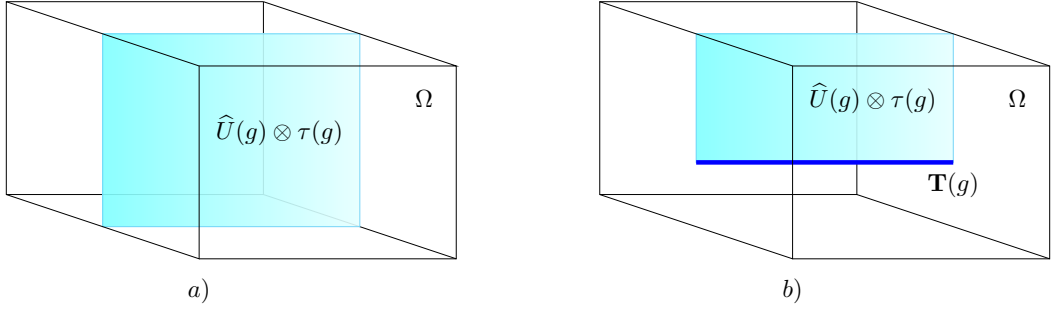
\begin{figure}
    \centering
    \scalebox{0.85}{
 \begin{tikzpicture}[baseline={(0,0.5)}]
\draw[] (0,0) -- (4,0) -- (4,3); \draw (4,0) -- (7,-1);
\draw[color=white!50!cyan, left color= white!50!cyan, right color=white!90!cyan, opacity=0.9] (1.5,-0.5) -- (1.5,2.5) -- (5.5,2.5) -- (5.5,-0.5) -- cycle;
\draw (0,0) -- (3,-1) -- (3,2) -- (0,3) -- cycle;    
 \draw (3,-1) -- (7,-1) -- (7,2); \draw (3,2) -- (7,2) -- (4,3); \draw (0,3) -- (4,3);
 \node at (6.5,1.5) {$\Omega$};
\node at (4.25,1) {$\widehat{U}(g) \otimes \tau(g)$};
\node at (3,-1.5) {$a)$};
\end{tikzpicture} 
} \qquad \qquad 
\scalebox{0.85}{
\begin{tikzpicture}[baseline={(0,0.5)}]
\draw[] (0,0) -- (4,0) -- (4,3); \draw (4,0) -- (7,-1);
\draw[color=white!50!cyan, left color= white!50!cyan, right color=white!90!cyan, opacity=0.9] (1.5,0.5) -- (1.5,2.5) -- (5.5,2.5) -- (5.5,0.5) -- cycle;
\draw[line width =2, blue] (1.5,0.5) -- (5.5,0.5);
\draw (0,0) -- (3,-1) -- (3,2) -- (0,3) -- cycle;    
\draw (3,-1) -- (7,-1) --(7,2) -- (4,3); \draw (3,2) -- (7,2); \draw (0,3) -- (4,3);
\node at (6.5,1.5) {$\Omega$};
\node at (4.25,1.5) {$\widehat{U}(g) \otimes \tau(g)$};
 \node at (5.5,0.15) {$\mathbf{T}(g)$};
 \node at (3,-1.5) {$b)$};
\end{tikzpicture} 
}
    \caption{a) An SPT in the presence of a topological defect $\widehat{U}(g)$, which is decorated by a lower dimensional SPT $\tau(g)$. b) Terminating the invertible interface to define a topological monodromy defect induces a decoration by a lower-dimensional TQFT $\mathbf{T}(g)$. On the mapping torus $Y^{d+1} = X^{d} \times S^1$ the left and right sides are identified.}
    \label{fig: inflow}
\end{figure}

Let us denote by $\widehat{U}(g)$ the $\bG$ symmetry defect in the SPT $\Omega$, in order to distinguish it from the symmetry defect $U(g)$ in the gapless setup. As discussed previously, its presence breaks the bulk symmetry to the centralizer $C_\bG(g)$ of $g$.
While completely transparent to local dynamics, $\widehat{U}(g)$ is not implemented trivially in the SPT phase: its worldvolume can be decorated by a lower dimensional invertible phase, which we denote $\tau(g)$. In the condensed matter literature, this is known as the decorated domain wall construction \cite{Chen:2013xva,Fidkowski:2016svr,Wang:2021nrp}. 
The decoration is not arbitrary and can be extracted from the SPT partition function in the presence of a fixed flat background creating the symmetry defect $\widehat{U}(g)$.
More precisely, one considers a mapping torus $Y^{d+1} = X^d \times S^1$ with a fixed holonomy equal to $g$ around the compact circle. This corresponds to a twisted compactification where the $\widehat{U}(g)$ defect is placed on $X^d$ at a given angular coordinate on the $S^1$. We have:
\be
Z_\Omega [X^d \times S^1; g] = Z_{\tau(g)}[X^d] \, ,
\ee
where $Z$ is the partition function
\be
Z_\Omega[Y] = \exp\left(i \int_Y A^*\Omega \right) \, ,
\ee 
$A$ is a background $\bG$ gauge field and ${}^*$ denotes the pullback.
The setup is depicted on the left of Figure \ref{fig: inflow}.
This compactification is described by a precise mathematical map, often referred to as transgression, or slant product:
\be
\tau : \mathbf{SPT}_{\bG}^{d+1} \longrightarrow \mathbf{SPT}_{C_{\bG}(g)}^{d} \, .
\ee
The map can be made more explicit by introducing a fixed flat reference background field 
\be 
{A}(g) = \text{PD}(\widehat{U}(g)) \, ,
\ee
where PD denotes the Poincare dual. A defect background field can be described by perturbing ${A}(g) \longrightarrow {A}(g) + B$, with $B$ a $C_\bG(g)$ gauge field. This assures that ${A}(g)$ is well defined under $B$ gauge transformations. 
For a discrete symmetry and a group-cohomology type SPT (a class $\Omega \in H^{d+1}(\bG,U(1))$), the slant product is given by the explicit formula:
\be
H^{d}(C_{\bG}(g), U(1)) \ni \tau(g) = \prod_{j=0}^d \Omega(h_1, ..., h_j, g, h_{j+1}, ..., h_d)^{(-1)^j} \, , \quad h_i \in C_{\bG}(g) \, .
\ee
On the other hand, for infinite order anomalies of continuous symmetries, the SPT $\Omega$ is described by a Chern-Simons term for the relevant symmetry, which is a function of $A$ and its field strength $F = dA + A\wedge A$. Computing a naive dimensional reduction on the mapping torus is an incorrect procedure, as the functional we are integrating is not gauge invariant. 
Rather, we identify:
\be
Z_{\text{CS}}[X^d \times S^1; g] = \exp\left( i \int_{Y^{d+2} = X^{d} \times D^2} P_{d+2}(F) \right) \equiv Z_{\tau(g)}[X^d] \, ,
\ee
where $P_{d+2}(F)$ is the anomaly polynomial.\footnote{ 
For example, a $U(1)^3$ anomaly has:
\be
P_{6}[F] = 2 \pi\frac{k}{6} \left( \frac{F}{2\pi}\right)^3 \, ,
\ee
where $k=1$ corresponds to a single (3+1)d Weyl fermion, and gives rise to well-defined (4+1)d Spin-SPT.}
Here we have chosen for simplicity a bulk topology $X^{d} \times D^2$, but the definition does not depend on the extension since $\int_{Y^{d+2}} P_{d+2} = 2 \pi \bZ$, for closed $Y^{d+2}$.
The presence of monodromy on the boundary of the disk forces:
\be
\exp\left(i \int_{D^2} F \right) = g \, ,
\ee
so we thread a fractional flux at the center of the disk.
The construction extends to higher-form symmetry, where it leads to a class:
\be
\tau(g) \in \mathbf{SPT}^{(d-p)}_{C_\bG(g)} \, . 
\ee
More abstractly, if anomalies are represented by the appropriate cobordism group, the transgression induces a pushforward operation on them. We will not need this in the present work, however.

We will actually need to be more careful than this, as we will soon consider an open defect $\widehat{U}(g)$. In this case, we will need to be mindful not only of $\tau(g)$ as a cohomology class, but, if trivial, we will also need to carefully track its trivialization, as it might not be possible to trivialize $\tau(g)$ while preserving the bulk symmetry. In such cases, $\tau(g)$ should be interpreted as the \emph{response} of its edge modes. 

\subsection{The topological monodromy $\mathbf{T}(g)$} 

Having discussed how to consistently decorate bulk symmetry defects in the bulk SPT, we now move to the question of their termination, which defines the topological monodromy defect $\mathbf{T}(g)$. We will use this to infer properties of the bulk SPT, and to then describe monodromy defects in gapless anomalous theories by inflow.

In the presence of a dynamical monodromy defect, the bulk defect $\widehat{U}(g)$ must also be terminated topologically, as seen on the right side of Figure \ref{fig: inflow}.
To preserve bulk topological invariance, the SPT $\tau(g)$ must thus be supplemented with a gapped boundary condition, $\mathbf{T}(g)$.

There is no unique choice of $\mathbf{T}(g)$. This follows e.g. from the ability of decorating it by a decoupled $(d-1)$-dimensional TQFT.
In practice, however, there is often a ``minimal" choice, which e.g. minimizes the partition function of $\mathbf{T}(g)$ on compact manifolds. The exact realization of a monodromy defect will depend on the microscopic details of the theory, but a set of its topological properties, which descend from the bulk SPT, will be universal. Depending on $\tau(g)$ we can encounter several distinct scenarios.

\paragraph{$\mathbf{T}(g)$ spontaneously breaks $\bG$.} Often, a gapped boundary condition $\mathbf{T}(g)$ exists, but it spontaneously breaks the $\bG$ (or, more precisely $C_\bG(g)$) symmetry, giving rise to topological edge modes. This situation arises when the transgression $\tau(g)$ is a nontrivial SPT that exhibits \emph{symmetry-enforced gaplessness}.
Several bulk realizations of this phenomenon are described, for example, in \cite{Cordova:2019bsd} and are enforced by a nontrivial integral of the original SPT phase over a mapping torus $M = S^{q} \times S^{d-q+1}$:
\be
\exp\left( i \int_{M} A^*\tau(g) \right) \neq 1 \, .
\ee
As the edge modes of the monodromy defect $\mon{g}$ must spontaneously break the residual $C_{\bG}(g)$ symmetry, the twisted Hilbert space becomes a direct sum:
\be
\cH_g = \bigoplus_{m} \cH_g(m) \, ,
\ee
where $\cH_g(m)$ corresponds to a vacuum of the topological order $\mathbf{T}(g)$.
A classic example appears when $\tau(k)$ is a discrete chiral anomaly for $\bG=\bZ_N$
\be
\tau(k) = \frac{2 \pi k}{N} \int A \beta(A) \, ,
\ee
this can originate e.g. from the (3+1)d SPT 
\be
\Omega(A, \widetilde{A}) = \frac{2 \pi}{N} \int \widetilde{A} \cup A \beta(A) \, ,
\ee
where $\widetilde{A}$ is a $\bZ_N$ gauge field compactified with holonomy $e^{2\pi i k/N}$.
Consequently, the topological monodromy defect $\mathbf{T}(k)$ is a (1+1)d TQFT characterized by a spontaneously broken $\bZ_N$ global symmetry, and decomposes into a direct sum on $N$ defects.
This is also a recurrent example in (2+1)d type III anomalies: in this case the bulk theory is a class $\Omega \in H^3(\bG, U(1))$ and $\tau(g) \in H^2(C_\bG(g),U(1))$ a nontrivial class. $\mathbf{T}(g)$ is a topological quantum mechanics whose Hilbert space transforms in a projective representation of $ C_\bG(g)$, which cannot be one-dimensional. The simplest example is $\bG = \bZ_2^3$ with anomaly $\Omega =  \pi \int A_1 A_2 A_3$ and $\tau(k_1) =  \pi \int A_2 A_3$, which is matched by a topological quantum mechanics carrying a projective representation of $\bZ_2 \times \bZ_2$.
    
\paragraph{$\mathbf{T}(g)$ transforms under an extension of $\bG$.} In this case, the gapped boundary condition $\mathbf{T}(g)$ exists \emph{and} fully preserves the $C_{\bG}(g)$ symmetry.
This can typically occur in two ways:

In several examples $\tau(g) = d \eta(g)$ is trivial, but the trivialization is not a well defined function on $C_{\bG}(g)$. Then $\mathbf{T}(g)$ is a symmetric, invertible defect, but the group acting on $\mathbf{T}(g)$ $C_\bG(g)$ is extended:\footnote{If instead $\eta(g)$ is a well-defined object, we can just tune local counterterms to cancel it by stacking its inverse.}
    \be
    1 \longrightarrow H \longrightarrow \Gamma(g) \longrightarrow C_\bG(g) \longrightarrow 1 \, .
    \ee
    The typical example are type I SPTs in (2+1)d:
    \be
    \Omega(A) = \frac{2 \pi}{N} \int A \beta(A) \, ,
    \ee
    where $\beta(A) = dA/N$ is the Bockstein map. In this case 
    $
    \tau(k) = \frac{2 \pi k}{N^2} \int d A $
  is trivialized by an ill-defined 1d Chern-Simons term:
    \be
    \mathbf{T}(k) = \frac{2 \pi k}{N^2} \int A \, .
    \ee
    The monodromy defect is an invertible object, which however transforms under an extension $\bZ_{N^2}$ of the original symmetry. The same mechanism can occur in higher dimensions, but the symmetry must extend to a higher-group $\Gamma(g)$:\footnote{This is in the spirit of the Wang-Wen-Witten mechanism \cite{Wang:2017loc}, which we further study below.}
    \be
    1 \longrightarrow H^{(1)} \longrightarrow \Gamma(g) \longrightarrow C_\bG(g) \longrightarrow 1 \, .
    \ee
    This is similar in spirit to the previous example due to symmetry fractionalization \cite{Wen:2001rto,Essin:2013vza,Barkeshli:2014cna,Chen:2014wse,Benini:2018reh,Brennan:2022tyl,Delmastro:2022pfo}: the fundamental excitations carry fractional charge, but are also charged under an emergent gauge symmetry.   
 A paradigmatic example stems from the chiral anomaly for $U(1)^3$. For simplicity consider a Spin anomaly (the more relevant case Spin$_c$ will be discussed at length in Section \ref{sec: chiralmon}) of the form
\be \label{eq: u1spinspt} \Omega = \frac{k}{24 \pi ^2} \int A (dA)^2 \, , \ \ \ \tau(\theta) = \frac{k \theta}{8 \pi^2} \int (dA)^2 \, , \ee
Where $k \in \bZ$.
Notice that the slant product provides a well defined map on Spin manifolds, as $\tau(2\pi) = \tau(0)$ by the quantization of the fermionic $\theta$ term. $\tau(\theta)$, for $\theta \neq \pi$, is a trivial SPT, as we can continuously dial $\theta$ to zero without breaking any symmetry. 
However, for fixed $\theta$, the trivialization cannot happen in a gapped and $U(1)$-symmetric way. To see this consider a ``domain-wall'' configuration:
\be
\theta(x_\perp) = 2 \pi p/N f(x_\perp) \, ,  \quad f(x_\perp \gg 1) = 1 \, , \quad f(x_\perp \ll 1) = 0 \, ,
\ee
Making the support of $f'$ arbitrarily narrow, we find:
\be \label{eq: wall}
\int \tau(\theta(x_\perp)) = \frac{p}{4 \pi N} \int_{X^{3}} A d A \, ,
\ee
which is an ill-quantized Spin Chern-Simons term. To properly define it, we must match the Hall conductivity $\sigma_H =  p /N$ by an appropriate (2+1)d TQFT, interpreting \eqref{eq: wall} as its fractional quantum Hall response. This can only be done by gapped phases which are long-range entangled (i.e. topologically ordered), and the symmetry is thus extended to a higher group on the edge.
Due to the extension by a higher-form symmetry (which can be non-invertible), the twisted Hilbert space carries the nontrivial topological order.

There are also many cases where a nontrivial $\tau(g)$ can be matched by a gapped boundary condition which does not break its symmetry. Most importantly, all group cohomology SPTs:
\be
\tau(k) \in H^{d}(C_\bG(g), U(1)) \, , \quad d \geq 4 \, ,
\ee
admit symmetry preserving gapped boundary conditions by the Wang-Wen-Witten mechanism \cite{Wang:2017loc,Tachikawa:2017gyf}. This extends to fermionic SPTs which take values in the appropriate super-cohomology \cite{Wan:2025ymd,Debray:2025kfg,Debray:2026sqw}.
Again the mechanism uses an emergent topological order, along with symmetry fractionalization, to trivialize the SPT response.

\paragraph{$\mathbf{T}(g)$ is electrically charged under $\bG$} 
In this case, the monodromy defect describes a new object in the bulk SPT, which can be detected by linking with $\widehat{U}(g)$. This happens e.g. with type II anomalies:
\be
\Omega(A,B) = 2 \pi \int A \cup B \, ,
\ee
with $A \in H^{p+1}(Y^{d+1},\bG_1)$, $ B \in H^{d-p}(Y^{d+1}, \bG_2)$, and the cup product providing a bilinear pairing $\chi: \bG_1 \times \bG_2 \to U(1)$. 
A monodromy defect $\mathbf{T}(g_1)$ for $\bG_1$ is charged under $\bG_2$ with charge determined by $\chi$. $\mathbf{T}(g_1)$ must be an invertible defect: its self-fusion $\mathbf{T}(g_1) \times \mathbf{T}(g_1)^\dagger$ carries no $\bG_2$ charge and no twist, and an SPT admits no non-trivial topological defect of this type --- it can only be the identity.
This is precisely the setting in which string order parameters traditionally appear: the invertible defect $\mathbf{T}(g)$ must condense
\be
\langle \mathbf{T}(g) \rangle \neq 0 \, ,
\ee
proving that the $\bG$ symmetry is broken in the twisted sector by the standard mechanism. Notice however that the twisted Hilbert space remains one-dimensional.

\paragraph{A gapped $\mathbf{T}(g)$ does not exist} The most extreme scenario is when the bulk gap closes in the presence of localized flux.
This can only happen in specific circumstances, as most SPT phases $\tau(g)$ (and certainly all group-cohomology ones \cite{Kitaev:2011dxc,Wang:2017loc}) admit a symmetry-preserving gapped boundary condition, or can be trivially gapped by symmetry breaking \cite{Cordova:2019bsd}.

Let us provide counter-examples. The first is to consider the (4+1)d $U(1)$ Spin-SPT phase \eqref{eq: u1spinspt}, but for \emph{irrational} $\theta$. In this case the anomaly cannot be matched by any finite topological order, and we must either resort to non-compact TQFT (which, however, have infinite number of ground states), or to a gapless $\mathbf{T}(g)$. 
Similarly, one can consider crystalline SPTs \cite{Thorngren:2016hdm,Else:2018eas} which are enriched by lattice translations. For example, if $\bG = \bZ^{d} \times U(1)$, a single lattice translation can stack a (2+1)d background Chern-Simons term (an integer Quantum Hall state). The resulting dislocation defects then must carry gapless modes. It is very important here that the translation symmetry is of infinite order, as this process is ill-defined for finite symmetries since the $U(1)_k$ SPT has infinite order.

\subsection{Spectral Pump and Anomaly in the space of defect couplings}\label{ssec: pump} An interesting signature of the anomaly arises when studying continuous symmetries with nontrivial $\pi_1(\bG)$, e.g. $\bG=U(1)$.
In this setup we can envision adiabatically turning on a localized flux for the $\bG$ background gauge field $A$ around the monodromy defect's worldvolume
\be
\exp\left( i \oint_C A(s) \right) = g(s) \, , \ \ \ g(0) = g(1) \, , \quad s \in [0,1]
\ee
describing a nontrivial class $\gamma \in \pi_1(\bG)$. While the transgression $\tau(g)$ must be invariant between the two endpoints, its representative can change:
\be
\tau(g(1)) =  \tau(g(0)) + d \omega( \gamma ) \, .
\ee
Where we have highlighted the dependence on the homotopy class $\gamma$ only. The fact that $\omega(\gamma)$ does not depend on small deformations of $\gamma$ follows from the discreteness of the space of SPTs. 

Following our previous discussion, we conclude that the monodromy defect $\mathbf{T}(g(1))$ is modified by stacking with an invertible phase $\omega(\gamma)$:
\be
\mathbf{T}(g(1)) = \mathbf{T}(g(0)) \otimes \omega(\gamma) \, .
\ee
For example, if $\mathbf{T}(\unit)$ is the trivial defect, the pump decorates it by the SPT $\omega(\gamma)$.
This is a monodromy-defect version of the Thouless pump \cite{Thouless:1983quantization}.
Similar phenomena appear in recent discussions of anomalies in the space of (defect) couplings \cite{Cordova:2019jnf,Inamura:2024jke,Komargodski:2025jbu,Choi:2025ebk,Copetti:2025sym}.
In this context, we should regard $\mathbf{T}(g)$ as a defect family over $\bG$. Notice that we do not need $\mathbf{T}(g)$ to be symmetric for the construction to hold: the spectral pump probes the homotopy of $\bG$ rather than $C_\bG(g)$.

The anomaly in the space of couplings also comes with another striking feature: since along the loop there is a nontrivial spectral flow, there must be a point at which level crossing happens. In higher dimension this usually means that we have an emergent symmetry $C$ (e.g. time reversal or charge conjugation) at a point $g^*$ on the loop.
A $C$-symmetric $\mathbf{T}(g^*)$ then must either break this symmetry spontaneously, or have the correct anomaly to absorb the pumped SPT $\omega(\gamma)$. So either:
\begin{itemize}
    \item $\mathbf{T}(g^*) = \mathbf{T}^+(g^*) \oplus \mathbf{T}^-(g^*)$, and:
    \be
    \mathbf{T}^-(g^*) = \mathbf{T}^+(g^*) \otimes \omega(\gamma) \, ,
    \ee
    and the two theories are exchanged by the emergent symmetry at $g^*$;
    \item or, $\mathbf{T}(g^*)$ is invariant under a further anomalous symmetry (C) under which:
    \be
    C \mathbf{T}(g^*) = \mathbf{T}(g^*) \otimes \omega(\gamma) \, ,
    \ee
    thus, it remains consistent by absorbing the pumped SPT. Notice that in this case the anomaly must descend from the bulk by transgression.
\end{itemize}
We will see examples of such level crossing in Section \ref{sec: chiralmon} and Appendix \ref{sec: 1p1}.
We now move onto the discussion of the main topic of this work: the consequences of the decorated domain wall construction for the physics of monodromy defects $\mon{g}$ in \emph{gapless} theories.

\section{Gapless theories and their monodromy defects}\label{sec: gapless}
After having reviewed the properties of monodromy defects in gapped SPT phases, we now move to the discussion of gapless theories.
The main tool to study gapless theories will be to realize them as the boundary of a bulk SPT, and to use the inflow mechanism to infer properties of their monodromy defects.

\subsection{Monodromy defects as domain walls}\label{sec: mondw}
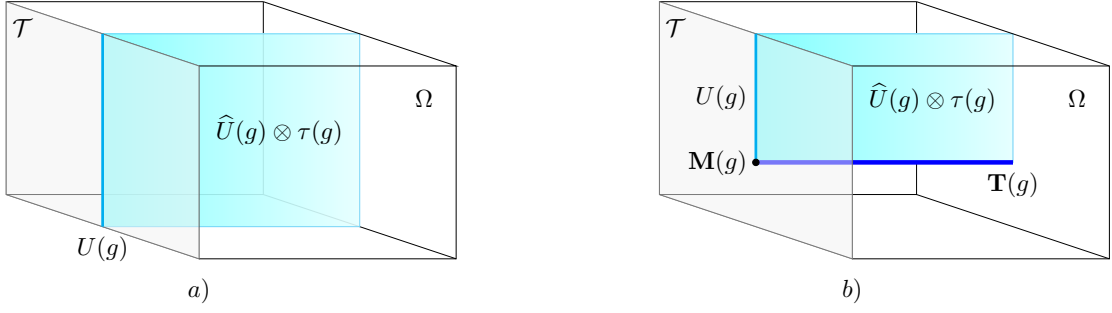
\begin{figure}
    \centering
    \scalebox{0.85}{
 \begin{tikzpicture}[baseline={(0,0.5)}]
\draw[] (0,0) -- (4,0) -- (4,3); \draw (4,0) -- (7,-1);
\draw[color=white!50!cyan, left color= white!50!cyan, right color=white!90!cyan, opacity=0.9] (1.5,-0.5) -- (1.5,2.5) -- (5.5,2.5) -- (5.5,-0.5) -- cycle;
\draw (0,0) -- (3,-1) -- (3,2) -- (0,3) -- cycle;    
 \draw (3,-1) -- (7,-1) -- (7,2); \draw (3,2) -- (7,2) -- (4,3); \draw (0,3) -- (4,3);
 \node at (6.5,1.5) {$\Omega$};
\node at (4.25,1) {$\widehat{U}(g) \otimes \tau(g)$};
\node at (3,-1.5) {$a)$};

\filldraw[color=gray!10, opacity=0.5] (0,0) -- (3,-1) -- (3,2) -- (0,3) -- cycle;
\draw[cyan, very thick] (1.5,-0.5) node[below,black] {$U(g)$} -- (1.5,2.5);
\node at (0.25,2.6) {$\cT$};

\end{tikzpicture} 
} \qquad \qquad \qquad
\scalebox{0.85}{
\begin{tikzpicture}[baseline={(0,0.5)}]
\draw[] (0,0) -- (4,0) -- (4,3); \draw (4,0) -- (7,-1);
\draw[color=white!50!cyan, left color= white!50!cyan, right color=white!90!cyan] (1.5,0.5) -- (1.5,2.5) -- (5.5,2.5) -- (5.5,0.5) -- cycle;
\draw[line width =2, blue] (1.5,0.5) -- (5.5,0.5);
\draw (0,0) -- (3,-1) -- (3,2) -- (0,3) -- cycle;    
\draw (3,-1) -- (7,-1) --(7,2) -- (4,3); \draw (3,2) -- (7,2); \draw (0,3) -- (4,3);
\node at (6.5,1.5) {$\Omega$};
\node at (4.25,1.5) {$\widehat{U}(g) \otimes \tau(g)$};
 \node at (5.5,0.15) {$\mathbf{T}(g)$};
 \node at (3,-1.5) {$b)$};

\filldraw[color=gray!10, opacity=0.5] (0,0) -- (3,-1) -- (3,2) -- (0,3) -- cycle;
\draw[cyan, very thick] (1.5,0.5) -- (1.5,2.5);
\node at (0.25,2.6) {$\cT$}; \node[left] at (1.5,1.5) {$U(g)$};
\draw[fill=black] (1.5,0.5) node[left] {$\mon{g}$} circle (0.05);

\end{tikzpicture} 
}
    \caption{The inflow setup to describe an anomalous monodromy defect $\mon{g}$ in a theory $\cT$. a) The boundary symmetry defect $U(g)$ extends to the decorated SPT defect $\widehat{U}(g)$. b) Terminating the invertible interface to define a monodromy defect shows that $\mon{g}$ is an interface between $U(g)$ and $\mathbf{T}(g)$.}
    \label{fig: inflow2}
\end{figure}

If the gapless theory $\cT$ has an anomalous global symmetry $\bG$, we can realize it at the boundary of an SPT phase $\Omega$ for $\bG$. The monodromy defect $\mon{g}$ of $\cT$ is then identified with the boundary of the topological monodromy $\mathbf{T}(g)$, itself a gapped boundary condition of the decorated symmetry defect $\widehat{U}(g) \otimes \tau(g)$. The setup is depicted on the right of Figure \ref{fig: inflow2}.
This perspective is powerful, as it allows to infer some of the properties of $\mon{g}$ from the inflow analysis of $\mathbf{T}(g)$, without resorting to a detailed analysis of the gapless theory $\cT$.
We can recover a purely boundary perspective by moving the defect $\mathbf{T}(g)$ to the boundary, and acting with $\widehat{U}(g) \otimes \tau(g)$ on the right-half space, thus stacking $\tau(g)$ on the right half of the boundary gapless theory.
Thus, the presence of an 't Hooft anomaly implies that the topological defect $U(g)$ acts as a domain wall between two different theories $\cT$ and $\cT'$, which differ by the stacking of the SPT $\tau(g)$:
\be
\cT' =  \cT  \otimes \tau(g)\, .
\ee
If $\tau(g)$ is nontrivial, we have two choices: we either accept to treat the symmetry defect as a domain wall between (mildly) different theories, or we insist on treating it as an interface between $\cT$ and itself, by introducing an appropriate topological decoration $\mathbf{T}(g)$ on $U(g)$. 
This construction has been used in \cite{Kaidi:2021xfk} to describe non-invertible symmetry defects in (3+1) dimensional systems. 

\begin{figure}[t!]
    \centering
    \begin{equation*}
    \scalebox{0.85}{
    \begin{tikzpicture}[baseline={(0,1)}, x={(1.4cm,-0.42cm)}, y={(0.7cm,0.4cm)}, z={(0cm,1.4cm)}]
        \filldraw[gray!10, opacity=0.7, draw=gray!80!black, thick, line join=round] plot[domain=0:180, samples=50] ({2*cos(\x)}, {2*sin(\x)}, 0) -- plot[domain=180:0, samples=50] ({2*cos(\x)}, {2*sin(\x)}, 2) -- cycle;
        \draw[gray!80!black, thick, opacity=0.7] ({2*cos(26.57)}, {2*sin(26.57)}, 0) -- ({2*cos(26.57)}, {2*sin(26.57)}, 2);
        
        \foreach \ang in {60, 90, 120, 150} {
            \draw[gray!50!black, thin, opacity=0.3] ({2*cos(\ang)}, {2*sin(\ang)}, 0) -- ({2*cos(\ang)}, {2*sin(\ang)}, 2);
        }
        \filldraw[gray!10, opacity=0.8, draw=gray!80!black, thick, line join=round] (0,0,2) -- plot[domain=0:180, samples=50] ({2*cos(\x)}, {2*sin(\x)}, 2) -- cycle;

        \filldraw[blue, opacity=0.5, draw=blue!80!black, thick, line join=round] (0,0,0) -- (-2,0,0) -- (-2,0,2) -- (0,0,2) -- cycle;
        \filldraw[blue!10, opacity=0.7, draw=blue!50!black, thick, line join=round] (0,0,0) -- (2,0,0) -- (2,0,2) -- (0,0,2) -- cycle;

        \draw[ultra thick, black] (0,0,0) -- (0,0,2);
        
        \node[black] at (1,0,1) {$U(g)$};
        \node[black] at (-1,0,1) {$\mathbf{T}(g)$};
        \node[below] at (0,0,0) {$\mon{g}$};
        \node[black] at (0,1,2) {$\tau(g)$};
    \end{tikzpicture}
    $\quad \simeq \quad$ 
    \begin{tikzpicture}[baseline={(0,1)}, x={(1.4cm,-0.42cm)}, y={(0.7cm,0.84cm)}, z={(0cm,1.4cm)}]
        \filldraw[blue, opacity=0.5, draw=blue!80!black, thick, line join=round] (0,0,0) -- ({2*cos(60)},{2*sin(60)},0) -- ({2*cos(60)},{2*sin(60)},2) -- (0,0,2) -- cycle;
        
        \filldraw[gray!10, opacity=0.7, draw=gray!80!black, thick, line join=round] plot[domain=0:60, samples=30] ({2*cos(\x)}, {2*sin(\x)}, 0) -- plot[domain=60:0, samples=30] ({2*cos(\x)}, {2*sin(\x)}, 2) -- cycle;
        
        \draw[gray!80!black, thick, opacity=0.7] ({2*cos(26.57)}, {2*sin(26.57)}, 0) -- ({2*cos(26.57)}, {2*sin(26.57)}, 2);
        
        \foreach \ang in {15, 45} {
            \draw[gray!50!black, thin, opacity=0.3] ({2*cos(\ang)}, {2*sin(\ang)}, 0) -- ({2*cos(\ang)}, {2*sin(\ang)}, 2);
        }

        \filldraw[gray!10, opacity=0.8, draw=gray!80!black, thick, line join=round] (0,0,2) -- plot[domain=0:60, samples=30] ({2*cos(\x)}, {2*sin(\x)}, 2) -- cycle;
        
        \filldraw[blue!10, opacity=0.7, draw=blue!50!black, thick, line join=round] (0,0,0) -- (2,0,0) -- (2,0,2) -- (0,0,2) -- cycle;

        \draw[ultra thick, black] (0,0,0) -- (0,0,2);
        
        \node[black] at (1,0,1) {$U(g)$};
        \node[black] at ({cos(60)}, {sin(60)}, 1) {$\mathbf{T}^\dagger(g)$};
        \node[below] at (0,0,0) {$\mon{g}$};
        \node[black] at ({cos(30)}, {sin(30)}, 2) {$\tau(g)$};
    \end{tikzpicture}
    $\quad \simeq \quad$
    \begin{tikzpicture}[baseline={(0,1)}, x={(1.4cm,-0.42cm)}, y={(0.7cm,0.84cm)}, z={(0cm,1.4cm)}]
        \filldraw[blue!15, opacity=0.7, draw=blue!60!black, thick, line join=round] (0,0,0) -- ({2*cos(26.57)},{2*sin(26.57)},0) -- ({2*cos(26.57)},{2*sin(26.57)},2) -- (0,0,2) -- cycle;
        
        \draw[ultra thick, black] (0,0,0) -- (0,0,2);
        
        \node[black] at ({1.0*cos(26.57)}, {1.2*sin(26.57)}, 1) {$U(g) \otimes \mathbf{T}^\dagger(g)$};
        \node[below] at (0,0,0) {$\mon{g}$};
    \end{tikzpicture}
}
 \end{equation*}
    \caption{Using the topological decoration $\mathbf{T}(g)$ to correctly define a twisted Hilbert space. The cyan region represents the stacking of $\tau(g)$; rotating $\mathbf{T}(g)$ around the junction $\mon{g}$ recasts the setup as a single composite defect $U(g) \otimes \mathbf{T}^\dagger(g)$.}
    \label{fig:monohilb}
\end{figure}
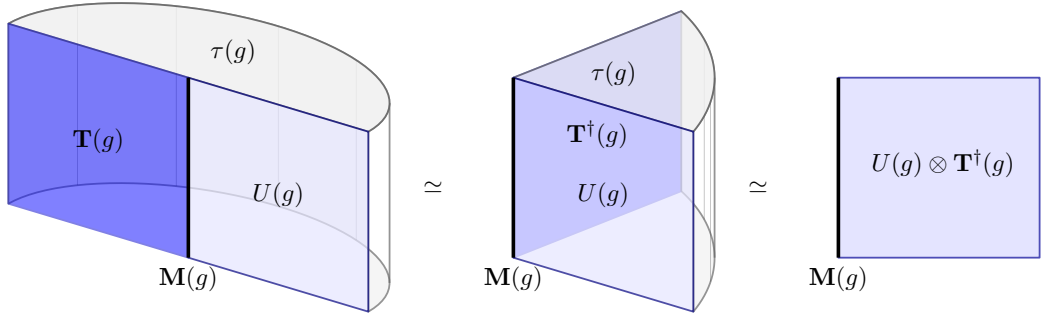

Thus, in the anomalous setting, the correct definition of $\mon{g}$ is as an interface between the symmetry defect $U(g)$ and a topological domain wall $\mathbf{T}(g)$. 
We can recover a more ``local'' description by rotating $U(g)$ around $\mon{g}$, as depicted in the last step of Figure \ref{fig:monohilb}, leading to a well-defined monodromy defect for $U(g) \otimes \mathbf{T}^\dagger(g)$.
With this object, we can arrive at a correct description of e.g. the twisted Hilbert space.
Thus, in the presence of an anomaly, we must give up on defining a twisted Hilbert space $\cH_g$, unless we choose a topological decoration $\mathbf{T}(g)$. The precise nature of the decoration (which might or might not exist) will depend again on the properties of $\tau(g)$. 

\subsection{Anomaly-induced properties of $\mon{g}$}
In Section \ref{sec: topomon} we have outlined how the bulk SPT phase $\Omega$ endows monodromy defects $\mathbf{T}(g)$ with nontrivial topological properties. Now let us outline how the four categories there translate to precise predictions regarding the properties of $\mon{g}$.

\paragraph{$\mathbf{T}(g)$ spontaneously breaks $\bG$} The topological monodromy defect $\mathbf{T}(g)$ has multiple vacua connected by the symmetry action. 
This implies that the twisted Hilbert space $\cH_g$ also shares the same property, and its energy levels are degenerate due to the broken symmetry. In other words, the defect $\mon{g}$ is \emph{not a simple defect}, and admits the decomposition:
\be
\mon{g} = \bigoplus_{m}  \mon{g}_m \, ,
\ee
where $m$ indicates a vacuum of $\mathbf{T}(g)$.
Examples will appear in Section \ref{sec: chiralmon} and Appendix \ref{sec: 1p1}.

Alternatively, states and operators within the twisted Hilbert space can be thought of as ``standard'' particles dressed by TQFT degrees of freedom. This allows us to neatly decouple the consequences of the anomaly from the underlying dynamics.

\paragraph{$\mathbf{T}(g)$ transforms under an extension of $\bG$} This is the most interesting scenario. Again, the physics takes two prominent forms:
\begin{itemize}
    \item The extension is by another 0-form symmetry, as in the case of the (1+1)d winding symmetry of the compact boson. This implies that defect operators carry fractional quantum numbers under the original $\bG$ symmetry.
    \item The extension involves a higher-form symmetry. Then the emergent higher-form symmetry operators in $\mathbf{T}(g)$ can terminate on $\mon{g}$, giving rise to a new sector of protected edge modes. Due to $\bG$ symmetry fractionalization, which is needed to match the anomaly, these modes carry charge under the bulk $\bG$ symmetry. The most interesting example, which we treat in Section \ref{sec: chiralmon}, is that of (3+1)d chiral symmetry. In this case the monodromy defect $\mon{\theta}$ is endowed with a novel sector of chiral protected edge modes.
\end{itemize}

\paragraph{Electrically charged $\mathbf{T}(g)$} 
When the topological monodromy defect $\mathbf{T}(g)$ is electrically charged under the $\bG$ symmetry, it must be identified with an electrically charged object in the gapless boundary theory.

In theories with 1-form symmetry this object is often the symmetry generator itself, leading to the well-known fact that defects charged under higher-form symmetries cannot be terminated in any way \cite{Rudelius:2020orz}. This can be understood as an empty twisted Hilbert space, or as the possibility of defining monodromy defects only as domain walls between likewise charged objects. A similar phenomenon takes place in (1+1)d adjoint QCD \cite{Komargodski:2020mxz}.

We will not consider this scenario further in this work.

\paragraph{No topological $\mathbf{T}(g)$} In this case, the monodromy defect is ill-defined, in the sense that it will not preserve the $SO(2)$ rotation symmetry, and is not a localized codimension-2 defect. The twisted Hilbert space is ill-defined, or empty. Concrete realizations of this scenario ( (4+1)d $U(1)$ Spin-SPTs at irrational $\theta$ and crystalline SPTs enriched by lattice translations ) were discussed at the end of Section \ref{sec: topomon}.

\paragraph{Consequences of the spectral pump} The spectral pump of Section \ref{ssec: pump} has interesting consequences for gapless defects. Suppose we start from a trivial monodromy defect $\mon{0}= \unit$. Pumping a unit of $\bG$ flux leads to the attachment of a $\bG$ SPT $\omega(\gamma)$. Thus, in the presence of flux, the trivial defect is dressed by a set of protected edge modes $\mathbf{N}$ of $\omega(\gamma)$:
\be
\mon{2 \pi} = \mathbf{N} \, .
\ee
This is a gapless manifestation of the Callan-Harvey inflow mechanism \cite{Callan:1984sa}.

\begin{figure}[t!]
    \centering
    \begin{tikzpicture}
        \draw[thick, mid arrow] (1.5,0) arc (0:180:1.5);
        \draw[thick, mid arrow] (-1.5,0) arc (180:360:1.5);
        
        \filldraw[black] (1.5,0) circle (2pt) node[left=4pt] {$0$};
        
        \node[right=4pt] at (1.5,-0) {$\mon{2\pi} = \mathbf{N}$};
        
        \filldraw[black] (-1.5,0) circle (2pt) node[right=4pt] {$\pi$};
                \node[left=4pt] at (-1.5,-0) {$\mon{\pi} = \mon{\pi}^+ \oplus \mon{\pi}^-$};
    \end{tikzpicture}
    \caption{The spectral pump along the parameter space of the monodromy defect. As the parameter $\theta$ winds around the circle, an SPT phase $\omega(\gamma)$ is pumped. At the symmetric point $\theta=\pi$ ($g = g^*$), the energy levels cross, indicating that the defect either becomes non-simple ($\mon{g^*}^+ \oplus \mon{g^*}^-$) or hosts protected edge modes.}
    \label{fig: levelcross}
\end{figure}
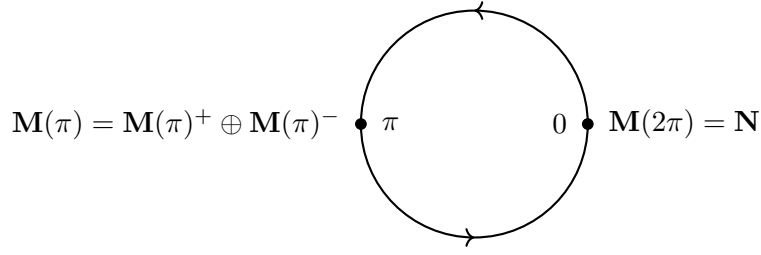

Furthermore, due to the nontriviality of the loop, we explicitly see the effect of level-crossing at $g = g^*$, see Figure \ref{fig: levelcross}. Either the monodromy defect is non-simple:
\be
\mon{g^*} = \mon{g^*}^+ \oplus \mon{g^*}^- \, ,
\ee
or it carries edge modes which soak up the mixed anomaly between the emergent $C$ symmetry and $C_{\bG}(g)$ described by $\omega(\gamma)$. In both cases this leads to new nontrivial physics on the $\mon{g^*}$ worldvolume.
Notice that, in general, the couplings on $\mon{g^*}$ can explicitly break the $C$-symmetry. In this case the anomaly matching argument is void.
Selected examples of the mechanism discussed in this Section can be found in Appendices \ref{sec: 1p1} and \ref{sec: 2Group}. We will now focus on the discussion of chiral monodromy defects in (3+1) dimensions, where these phenomena also appear.

\section{Chiral Monodromy Defects}\label{sec: chiralmon}
We now describe the effects of bulk 't Hooft anomalies on chiral monodromy defects. We will then apply these results to the case of a free Dirac fermion in (3+1)d.

The discussion is different depending on whether we consider bosons or fermions, and whether the fermions have even or odd $U(1)$ charge. 
In the examples we will consider bulk theories of fermions, in which case the $U(1)$ symmetry contains fermion parity $(-)^F$ as a $\bZ_2$ subgroup (thus, the charge of the elemental fermions is odd). 
This endows the theory with a Spin$_c$ structure:
\be
\int \frac{F}{2 \pi} = w_2 \, \mod 1 \, .
\ee
Physically, this expresses the spin-charge relation:
\be
2 s = Q \mod 1 \, .
\ee
In this case, the anomaly polynomial must take the form:
\be
I_6 = 2 \pi \left[ k \left( \frac{1}{6}\left(\frac{F}{2\pi}\right)^3 + \frac{F}{2\pi} \wedge \widehat{A}(R) \right) + 24 \widetilde{k_g}\frac{F}{2\pi} \wedge \widehat{A}(R) \right]   \, ,   \qquad k\, , \widetilde{k_g} \in \bZ \, ,
\ee
where $\widehat{A}(R) = \frac{1}{192 \pi^2} \Tr R \wedge R$ is the A-roof genus whose integral on a 4-manifold is $\int_{X_4} \widehat{A}(R) = -\frac{1}{8} \sigma$, with the signature $\sigma$ an integer. We will also use $k_g = k + 24\widetilde{k_g}$ to denote the total mixed anomaly. 
For a single Weyl fermion of (odd) charge $q$ we have $k = q^3$ and $k_g = k + 24\widetilde{k_g} = q$ and $q^3 - q = q(q+1)(q-1) \in 24 \bZ$. An even charge is incompatible with the Spin$_c$ structure.
For a spin theory instead $k$ and $k_g$ are independent integers, as the signature is a multiple of $16$.

We will also consider a Dirac fermion, which contains two Weyl components of charge $(1,1)$ and $(1,-1)$ under the $U(1)_V \times U(1)_A/\bZ_2^F$ symmetry. In this case the anomaly polynomial reads:
\be
I_6^{\text{Dirac}} = 2 \pi \left[ k \left( \frac{1}{3}\left(\frac{F_A}{2\pi}\right)^3 + \frac{F_A}{2\pi} \left(\frac{F_V}{2\pi}\right)^2 + 2\frac{F_A}{2\pi} \wedge \widehat{A}(R) \right) + 48 \widetilde{k_g}\frac{F_A}{2\pi} \wedge \widehat{A}(R) \right]   \, .
\ee
where again we use $k=q^3$ and $k_g = q$.

We will mostly consider the theory of a (3+1)d bulk Dirac fermion $\Psi$, which we take to be massless:
\be
S_{\text{bulk}} = i \int d^4 x \overline{\Psi} \gamma^\mu \partial_\mu \Psi \, .
\ee
The $[U(1)_V \times U(1)_A]/\bZ_2$ symmetry acts on the Dirac field as:
\bea
U(1)_V: &&\Psi \to e^{ 2 \pi i \mu} \Psi \, , \\
U(1)_A: &&\Psi \to e^{2 \pi i \nu \gamma_5} \Psi \, .
\eea
A chiral monodromy defect is sourced by localized axial magnetic flux:
\be
\frac{F_A}{2 \pi} = \nu \delta(\Sigma) \, .
\ee
Free fermion monodromy defects have been previously studied in e.g. \cite{Bianchi:2021snj} in the case of a vector $U(1)$ symmetry. In this case the authors observed nontrivial multi-valuedness of the defect b-function. 
As the Weyl fermions $\psi_L$ and $\psi_R$ remain decoupled in the presence of the axial flux, we can immediately generalize their results to our case.
\subsection{Transgression and topological dressing}
We now describe how to define monodromy defects for chiral fermions.
\paragraph{Weyl fermion} From the form of the 't Hooft anomaly of the single Weyl fermion, one can immediately compute the transgression, where we set $\theta = 2 \pi \nu$.
\be
\tau(\nu) = \nu \left[ k \left( \frac{1}{4 \pi} d A \wedge d A + \widehat{A}(R) \right) + 24 \widetilde{k_g} \widehat{A}(R) \right] = \nu \left[ \frac{k}{4 \pi} d A \wedge d A + k_g \widehat{A}(R) \right] \, .
\ee
The term between parenthesis is a well-quantized $\theta$ angle for the Spin$_c$ structure, see Appendix \ref{app: thetaperiod}.
For $\nu \neq 1/2$ this is a trivial SPT, which cannot however be trivialized by a well-quantized topological term. For rational $\nu = p/N \neq 1/2$ a gapped boundary condition is a topologically ordered fractional quantum hall state, whose Hall response must satisfy:
\be
\sigma_H = \frac{p}{N} k \, .
\ee
Furthermore, its chiral central charge is given by the coefficient of the $\widehat{A}$ term:
\be
c_- = \frac{p}{N} k_g \, .
\ee
The idea is now to match both by a (2+1)d topological order. In the Spin$_c$ case, we are free to stack integer quantum Hall states, but subject to the selection rule:
\be
\sigma_H - c_- = 0 \mod 8 \, , \ \ \ \ \text{for invertible states} \, .
\ee
This is to be contrasted with the Spin picture, in we can always offset the chiral central charge by half units.

Consider now $q=1$ which implies that $k=1, \widetilde{k_g}=0$, the strategy is valid in general. For rational $\nu = p/N$ we start by matching the Hall conductance by a minimal $\bZ_N$ TQFT $\cA_{N,p}$ \cite{Hsin:2018vcg}. See Appendix \ref{app: bostospin} for details about these theories and their compatibility with the Spin$_c$ structure.
This is done by symmetry fractionalization: the minimal theory carries a 1-form symmetry anomaly:
\be
\Omega_{m,N}(B) = \frac{ \pi m}{ N} \fP(B) \, ,
\ee
where $\fP$ is the Pontryagin square operation.\footnote{This is a refinement of the quadratic form $B \cup B$, see e.g. \cite{Benini:2018reh}.} 
In our conventions the generating line $L$ has spin
\be
\theta_L = \exp\left( \frac{\pi i p}{N} \right) \, ,
\ee
and charge $p$ under the one-form symmetry.
The bulk Hall response is matched by imposing the symmetry fractionalization condition \cite{Barkeshli:2014cna,Brennan:2022tyl,Delmastro:2022pfo,Brennan:2025acl}
\be
B = \kappa \frac{F}{2 \pi} \, .
\ee
Expanding the Pontryagin square this leads to the Hall response:
\be
\tau(\nu) = \frac{p \kappa^2}{4 \pi N} \int F \wedge F \, ,
\ee
which gives a Hall conductivity $\sigma_H = p \kappa^2/N$.
Physically, this means that a symmetry line carrying one-form symmetry charge $q$, which is attached to a background surface:
\be
\exp\left(\frac{2 \pi i q}{N} \int B \right) \, ,
\ee
is the worldline of a particle with fractional electric charge $\kappa q/N$. In this way, the symmetry group realized on anyons is effectively an $N$-fold cover of $U(1)$, which makes the fractional Chern-Simons level well-defined. Notice that $U(1)$ can only be fractionalized with invertible anyons \cite{Barkeshli:2014cna}, so the minimal block is universal and must be present in all the choices of $\mathbf{T}(\nu)$.

For $Np$ odd, $\cA_{N,p}$ theory is a Spin$_c$ theory only if $\kappa$ is also odd, as the fermion $\psi = L^N$ has charge $p \kappa$, which must be an odd integer.  
When $Np$ instead is even the theory is bosonic. The boson $L^N$ has charge $p \kappa$, and thus supports a Spin$_c$ structure only for even $p$. For odd $p$ and even $N$, we can still get the correct Hall response by considering $\cA_{4N, p}$, but with fractionalization class $\kappa = 2$. This is a Spin$_c$ theory as the boson has even charge.

The chiral central charge of these theories is however always an (half-) integer, thus, it cannot match the gravitational response for generic $p/N$. To solve this, we need to stack a non abelian topological order $\mathbf{T}_0$ such that:\footnote{See also \cite{Putrov:2023jqi} for a related construction.}
\be
c_-(\mathbf{T}_0) = \frac{p}{N} - c_-(\cA_{N,p}) \, .
\ee
This process can be rather messy, but it is generally solvable by an appropriate RCFT. 
We summarize a few examples in Table \ref{tab: ratnu}. In each case the matching takes the form $\mathbf{T}(\nu) = \cA_{N,p}^{\kappa} \otimes \mathbf{T}_0$, where $\cA_{N,p}^{\kappa}$ is the minimal $\bZ_N$ piece matching the Hall response and $\mathbf{T}_0$ is an extra factor compensating the chiral central charge.

\begin{table}[t]
\centering
\renewcommand{\arraystretch}{1.3}
\begin{tabular}{c|c|c|c}
$\nu$ & $\cA_{N,p}^{\kappa}$ & $c_{N,p}$ & $\mathbf{T}_0$ \\\hline
$1/3$  & $\cA_{3,1}^{\kappa=1}$  & $1$ & $SU(2)_4 \otimes \overline{SU(2)_{16}}$ \\
$2/3$  & $\cA_{3,2}^{\kappa=1}$  & $2$ & $U(1)_{2n} \otimes \overline{SU(2)_7}$ \quad ($n \in \bZ$) \\
$1/4$  & $\cA_{16,1}^{\kappa=2}$ & $1$ & $SU(2)_2 \otimes \overline{SU(2)_6}$ \\
$3/10$ & $\cA_{40,3}^{\kappa=2}$ & $-1$ & $U(1)_{2n}^2 \otimes \overline{\text{3Ising}}$ \\
\end{tabular}
\caption{Sample descriptions of $\mathbf{T}(\nu) = \cA_{N,p}^{\kappa} \otimes \mathbf{T}_0$ for a single Weyl fermion at rational $\nu$. The $\cA_{N,p}^{\kappa}$ factor matches the Hall response $\sigma_H = \nu$; the chiral central charge $c_{N,p}$ of the minimal piece is shown explicitly, with $\mathbf{T}_0$ chosen to compensate the residual $c_- = \nu - c_{N,p}$. 3Ising denotes the MTC of the tricritical Ising model.}
\label{tab: ratnu}
\end{table}
We conclude that, at these rational points, the monodromy defect $\mon{\nu}$ must be dressed by a nontrivial topological order $\mathbf{T}(\nu)$ and carries a sector of gapless chiral protected edge modes.

\paragraph{The time-reversal-invariant point $\nu=1/2$.}
At $\nu = 1/2$ the identification $\nu \to -\nu$ is preserved by time reversal $T$, enhancing the symmetry on the monodromy defect to
\be
\bG_{\nu = 1/2} = \big( U(1) \rtimes T \big)/ \bZ_2^F = \text{Pin}^+_c \, .
\ee
The transgressed response is the unique nontrivial Pin$_+^c$ SPT in $(3+1)d$:
\be \label{eq: tau-half}
\tau(1/2) = \frac{1}{8 \pi} F \wedge F + \frac{1}{2} \widehat{A}(R) \, .
\ee
By the spectral-pump argument of Section \ref{ssec: axialpump}, $\mon{1/2}$ sits at the level-crossing point of the loop $\nu \to \nu+1$ and must either break $T$ spontaneously or carry a $T$-anomalous topological order:
\begin{enumerate}
\item \emph{$T$-symmetry breaking.} The decoration splits into two vacua,
\be
\mathbf{T}(1/2) = \mathbf{T}^+ \oplus \mathbf{T}^- \, ,
\ee
related by a $U(1)_1$ response, corresponding to the two massive phases of a $(2{+}1)d$ Dirac fermion with positive/negative mass. In this realization $\mon{1/2}$ is not simple.
\item \emph{Topological order.} A symmetric $\mathbf{T}(1/2)$ is provided by the T-Pfaffian \cite{Bonderson:2013laa,Wang:2013gok,Chen:2013aha,Metlitski:2013uqa,Fidkowski:2013jua}:
\be
\text{T-Pfaffian} = \left( U(1)_8^{\kappa=2} \otimes \overline{\text{Ising}} \right)/\bZ_2 \, ,
\ee 
where $\kappa = 2$ is the choice of fractionalization class in the $U(1)_8$ theory. This topological order absorbs the Pin$_+^c$ anomaly while leaving $\mon{1/2}$ simple, with a universal chiral edge sector.
\end{enumerate}
We will see that the $T$-breaking option is realized by the free fermion system.



\paragraph{Irrational $\nu$} Finally, it is an open question how the transgression is matched when $\nu$ is irrational. In this case, our construction implies that $\mathbf{T}(\nu)$ must match an irrational Hall response, which is not realizable by any finite TQFT.
As explained previously, we would need to consider either a noncompact TQFT, such as those studied in \cite{GarciaEtxebarria:2022jky,Arbalestrier:2024oqg, Argurio:2024ewp}, or a gapless domain-wall. We have found a similar structure in the case of a (1+1) dimensional compact boson, where the symmetry acting on the monodromy defect, for irrational $\nu$, is $\bR$ rather than $U(1)$. The situation here is however more severe, and it is likely that a standard type of monodromy defect cannot be defined in this scenario.

\subsection{Axial pump} \label{ssec: axialpump}

We now describe the pump induced by inserting a $2\pi$ axial flux into the thin solenoid representing the monodromy defect, $\nu \to \nu + 1$. Following the general discussion of Section \ref{ssec: pump}, this loop traces a nontrivial element of $\pi_1(U(1)_A) = \bZ$ and pumps an invertible phase $\omega(\gamma)$ onto the worldvolume of $\mon{\nu}$.

For a single Weyl fermion the pumped SPT is a $U(1)_1$ phase:
\be \label{eq: pumped-spt}
\omega(\gamma) = \frac{1}{4 \pi} \int A_V \, d A_V  + \text{CS}_g \, .
\ee
For a Dirac fermion, the effect of the chiral transformation stacks between the $L$ and $R$ Weyl components and the pumped SPT is doubled, $2\cdot \omega(\gamma)$. By anomaly inflow, this requires surplus chiral edge modes localized on $\mon{2\pi}$: a single chiral $(1+1)d$ Weyl in the Weyl-fermion case, and two $(1+1)d$ Weyl fermions of the same chirality in the Dirac case.

This is the Callan--Harvey inflow mechanism \cite{Callan:1984sa} applied to a gapless bulk: the bound modes are the genuine $(1+1)d$ Weyl edge modes of the pumped invertible $U(1)_1$ SPT, and are resilient --- they cannot be removed without breaking the anomalous axial symmetry explicitly.

We will show below, by direct analysis of the multi-valuedness of the b-function (Section \ref{ssec: bfunc}), that for most families of monodromy defects these bound modes do not correspond to a decoupled $(1+1)d$ CFT, but to an interacting defect dressed by the topological orders identified above.

The axial pump also organizes the physics at the special point $\nu = 1/2$: this is the level-crossing point along the loop $\nu \in [0,1)$, where the general argument of Section \ref{ssec: pump} predicts that the monodromy defect must either spontaneously break the emergent time-reversal symmetry as $\mon{1/2}^+ \oplus \mon{1/2}^-$, or carry a topological order absorbing half of the pumped SPT. We worked this out explicitly in Section \ref{sec: topomon}; the free fermion analysis of Section \ref{sec: chiralmon} will confirm that the first option (symmetry breaking) is the one realized.

\subsection{A free fermion analysis} To better understand the physics of the monodromy defect, we can directly analyze the mode expansion of the free fermion in the background of a localized axial magnetic flux. Related studies have been performed in the past, see e.g. \cite{Bianchi:2021snj} for the vector case.

Throughout this analysis we distinguish two kinds of modes localized at the defect. \emph{Singular modes} diverge as $r^{-\alpha}$ at the location of the defect with $0 < \alpha < 1$ and are normalizable. \emph{Bound modes} correspond to the limiting case $\alpha = 1$ and are the genuine $(1+1)d$ Weyl fermions localized on the defect's worldvolume. 

It is clear that the left and right modes (according to $\gamma_5$) are not mixed in the computation and can be treated separately.
We will denote the (1+1)d fermionic modes appearing in the bulk-to-defect OPE by:
\be
\Psi = (\psi_L , \, \psi_R) \, , \quad \psi_L = (\chi_L^+, \chi_L^-) \, , \quad \psi_R = (\chi_R^-, \, \chi_R^+) \, .
\ee
where $L,R$ identify the 4d chirality (i.e. the eigenvalue of $\gamma^5$) while $\pm$ refers to the handedness of these fermions under the (1+1)d chiral symmetry localized on the defect. We choose the following basis of $\gamma$ matrices:
\be
\gamma^0 = \begin{pmatrix}
    0 & 1 \\
    1 & 0 
\end{pmatrix} \, , \quad \gamma^1 = \begin{pmatrix}
    0 & Z \\
    -Z & 0 
\end{pmatrix} \, , \quad \gamma^2 = \begin{pmatrix}
    0 & X \\
    -X & 0 
\end{pmatrix} \, , \quad \gamma^3 = \begin{pmatrix}
    0 & Y \\
    -Y & 0 
\end{pmatrix} \, .
\ee
With this choice of basis both $\gamma^5 = i \gamma^0 \gamma^1 \gamma^2 \gamma^3 = \text{diag}(\unit,-\unit)$ and the (1+1)d chirality matrix $\gamma_0 \gamma_1$ are diagonal. Notice that the 4d chirality eigenvalue flips the meaning of the 2d chirality.

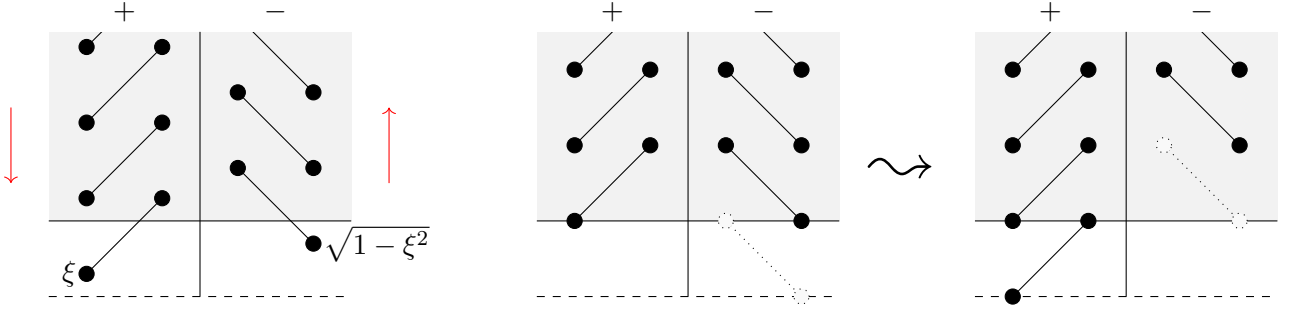
\begin{figure} 
    \begin{center}
    \begin{tikzpicture}[baseline={(0,1.5)}]
\fill[gray!10] (-2,1) rectangle (2,3.5);
\draw (-2,1) -- (2,1);
\draw[dashed] (-2,0) -- (2,0);

  \foreach \x in {0,...,3} {
        \draw (-1.5 , 0.3 + \x) -- (-0.5, 0.3 + \x + 1);
        \draw[fill=black] (-1.5 ,0.3 + \x) circle (0.1);
        \draw[fill=black] (-0.5, 0.3 + \x + 1) circle (0.1);
    }
    \node[left] at (-1.5,0.3) {$\xi$};
      \foreach \x in {0,...,2} {
        \draw (1.5 , 0.7 + \x) -- (0.5, 0.7 + \x + 1);
        \draw[fill=black] (1.5 ,0.7 + \x) circle (0.1);
        \draw[fill=black] (0.5, 0.7 + \x + 1) circle (0.1);
    }
        \node[right] at (1.5,0.7) {$\sqrt{1 - \xi^2}$};

\fill[white] (-2,3.5) rectangle (2,4.5);
\draw (0,0) -- (0,3.5); 
\node[above] at (-1, 3.5) {$+$};
\node[above] at (1, 3.5) {$-$};

\draw[red,->] (-2.5,2.5) --(-2.5,1.5);
\draw[red,<-] (2.5,2.5) --(2.5,1.5);
\end{tikzpicture}
\qquad \quad 
 \begin{tikzpicture}[baseline={(0,1.5)}]
\fill[gray!10] (-2,1) rectangle (2,3.5);
\draw (-2,1) -- (2,1);
\draw[dashed] (-2,0) -- (2,0);

  \foreach \x in {0,...,2} {
        \draw (-1.5 , 1 + \x) -- (-0.5, 1 + \x + 1);
        \draw[fill=black] (-1.5 ,1 + \x) circle (0.1);
        \draw[fill=black] (-0.5, 1 + \x + 1) circle (0.1);
    }
    
      \foreach \x in {1,...,3} {
        \draw (1.5 ,  + \x) -- (0.5,  + \x + 1);
        \draw[fill=black] (1.5 , + \x) circle (0.1);
        \draw[fill=black] (0.5,  + \x + 1) circle (0.1);
    }
    \draw[dotted] (0.5,1) -- (1.5,0);
    \draw[fill=gray!5, dotted] (0.5,1) circle (0.1);
    \draw[fill=gray!5, dotted] (1.5,0) circle (0.1);

\fill[white] (-2,3.5) rectangle (2,4.5);
\draw (0,0) -- (0,3.5); 
\node[above] at (-1, 3.5) {$+$};
\node[above] at (1, 3.5) {$-$};

\end{tikzpicture}
\ \ {\Huge $\leadsto$ } \ 
 \begin{tikzpicture}[baseline={(0,1.5)}]
\fill[gray!10] (-2,1) rectangle (2,3.5);
\draw (-2,1) -- (2,1);
\draw[dashed] (-2,0) -- (2,0);

  \foreach \x in {0,...,3} {
        \draw (-1.5 ,  + \x) -- (-0.5,  + \x + 1);
        \draw[fill=black] (-1.5 , + \x) circle (0.1);
        \draw[fill=black] (-0.5,  + \x + 1) circle (0.1);
    }
    \begin{scope}[shift={(0,1)}]
      \foreach \x in {1,...,2} {
        \draw (1.5 ,  + \x) -- (0.5,  + \x + 1);
        \draw[fill=black] (1.5 , + \x) circle (0.1);
        \draw[fill=black] (0.5,  + \x + 1) circle (0.1);
    }
    \draw[dotted] (0.5,1) -- (1.5,0);
    \draw[fill=gray!5, dotted] (0.5,1) circle (0.1);
    \draw[fill=gray!5, dotted] (1.5,0) circle (0.1);
    \end{scope}

\fill[white] (-2,3.5) rectangle (2,4.5);
\draw (0,0) -- (0,3.5); 
\node[above] at (-1, 3.5) {$+$};
\node[above] at (1, 3.5) {$-$};

\end{tikzpicture}
\end{center}

\caption{Left, the spectrum in the presence of a monodromy defect. The white area contains the singular zero modes, while gray areas are regular modes. The left and right sides are different 2d chiralities. The dashed line denotes the limit scaling $\sim r^{-1}$. Middle, the trivial defect. Right, spectral flow of the trivial defect: a left-moving zero mode is created at $\nu \to 1$.}
\label{fig: spectralweyl}
\end{figure}

Let us first understand the spectrum of a bulk Weyl fermion $\psi$ in the monodromy background. 
For simplicity, we will work on $\bR^{1,3}$ in cylindrical coordinates:
\be \label{eq: cylcol}
ds^2 = -dt^2 + dx^2 + dr^2 + r^2 d \theta^2 \, ,
\ee
with $\mon{g}$ on the $(t,x)$ plane. Boundary conditions around $\theta$ are twisted:
\be
\psi(t,x,r,\theta + 2 \pi) = -e^{-2\pi i \nu} \psi(t,x,r,\theta) \, .
\ee
The minus sign corresponds to standard anti-periodic spin structure on the cycle surrounding the defect and we pick $\nu \in [0,1)$.
In our basis, the Weyl components correspond to eigenmodes of the 2d chirality, which we indicate by $\pm$: $\psi = (\psi_+, \psi_-)$. We expand the Weyl fermion in modes:
\begin{equation}
    \psi(t,x,r,\theta) = e^{-i E t} e^{i p x} e^{i m \theta} \frac{1}{\sqrt{r}} \begin{pmatrix} f_+(r) \\ f_-(r) \end{pmatrix} \, ,
\end{equation}
where $m = n - \nu + 1/2$, for integer $n$. The $1/2$ shift takes care of the spin structure.
 The equations of motion, after some manipulation, take the form
\begin{align}
   &f_+' - \frac{m}{r} f_+ = - i \left(E-p \right) f_- \, , \\
   &f_-' + \frac{m}{r} f_- = -i \left( E + p \right)f_+ \, .
\end{align}
These can be massaged into a Bessel equation:
\be
f_+'' + \left(k^2 - \frac{(m-1)m}{r^2} \right) f_+ = 0 \, , \qquad k^2 = E^2 - p^2 \, .
\ee
A similar equation holds for the right movers, with $m \to - m$. We want solutions which are regular at $r=\infty$ and square-integrable. The solution takes the form:
\be
\begin{aligned}
\text{For } n>0: \quad & (f_+, f_-) = C \sqrt{r} \left( J_{m-1/2}(k r), -i \frac{k}{E-p} J_{m+1/2}(k r) \right) \, , \\
\text{For } n<0: \quad & (f_+, f_-) = \tilde{C} \sqrt{r} \left( J_{1/2-m}(k r), i \frac{k}{E-p} J_{-m-1/2}(k r) \right) \, .
\end{aligned}
\ee
In our chosen range for $\nu$, the first solution is the only available one when $n>0$, while the second only exists for $n<0$. Notice that, in the two cases, the left/right component of the spinor fixes the subleading mode uniquely. Both modes are regular at the origin, so no localized mode is present at this level.
For $n=0$, both solutions are valid and give rise to defect zero modes $\chi_{\pm}$, which behave asymptotically as
\bea
&\chi_+ : \, \psi_+ \sim r^{-\nu} \, , \quad && \psi_- \sim r^{-\nu +1} \, , \\
&\chi_- : \, \psi_- \sim r^{-1+\nu} \, , \quad && \psi_+ \sim r^{\nu} \, .
\eea
The physical interpretation is that $\chi_+$ (resp. $\chi_-$) is a singular chiral left- (resp. right-) moving $(1+1)d$ zero mode localized at the monodromy defect. Introducing a mixing parameter $\xi$, the bulk fermion has the small $r$ expansion:
\be
\psi \sim \xi \, r^{-\nu} \chi_+ + \sqrt{1 - \xi^2}\,  r^{-1 + \nu} \chi_- + \text{regular} \, .
\ee
Different $\xi$ correspond to different type of monodromy defects. A special choice is the one which reduces to the trivial defect as $\nu \to 0$, and corresponds to $\xi =1$. 
The next step is to discuss the spectral flow on the defect as we change $\nu$. We represent the ``sea'' of defect modes as in Figure \ref{fig: spectralweyl}.
As can be seen from the figure, the spectrum at $\xi=1$, $\nu=0$ is perfectly left-right symmetric (as it should be, since this is the trivial defect). As we flow to $\nu \to 1$ we create a left-moving zero mode at the defect, with no compensating right-moving counterpart. We will soon confirm that our anomaly arguments, combined with the study of the defect conformal anomaly (the b-function \cite{Jensen:2015swa,Wang:2020xkc}) exactly predict $\Delta c_L = 1$, $\Delta c_R = 0$.

\begin{figure}[t!] 
    \begin{center}
    \begin{tikzpicture}[baseline={(0,0)}]
    \fill[gray!10] (-2,1) rectangle (2,3.5);
    \draw (-2,1) -- (2,1);
    \draw[dashed] (-2,0) -- (2,0);
    
      \foreach \x in {0,...,3} {
            \draw (-1.5 , 0.25 + \x) -- (-0.5, 0.25 + \x + 1);
            \draw[fill=black] (-1.5 ,0.25 + \x) circle (0.1);
            \draw[fill=black] (-0.5, 0.25 + \x + 1) circle (0.1);
        }
        \node[left] at (-1.5,0.25) {$\xi_L$};
          \foreach \x in {0,...,2} {
            \draw (1.5 , 0.75 + \x) -- (0.5, 0.75 + \x + 1);
            \draw[fill=black] (1.5 ,0.75 + \x) circle (0.1);
            \draw[fill=black] (0.5, 0.75 + \x + 1) circle (0.1);
        }
            \node[right] at (1.5,0.75) {$\sqrt{1 - \xi_L^2}$};
    
    \fill[white] (-2,3.5) rectangle (2,4.5);
    \draw (0,0) -- (0,3.5); 
\node[above] at (-1, 3.5) {$+$};
\node[above] at (1, 3.5) {$-$};
    \draw[red,->] (-2.5,2.5) --(-2.5,1.5);
\draw[red,<-] (2.5,2.5) --(2.5,1.5);
\draw[blue,->] (-3,2.5) --(-3,1.5);
\draw[blue,<-] (3,2.5) --(3,1.5);
    \end{tikzpicture}
    \qquad 
    \begin{tikzpicture}[baseline={(0,0)}]
    \fill[gray!10] (-2,1) rectangle (2,3.5);
    \draw (-2,1) -- (2,1);
    \draw[dashed] (-2,0) -- (2,0);
    
      \foreach \x in {0,...,2} {
            \draw (-1.5 , 0.7 + \x) -- (-0.5, 0.7 + \x + 1);
            \draw[fill=black] (-1.5 ,0.7 + \x) circle (0.1);
            \draw[fill=black] (-0.5, 0.7 + \x + 1) circle (0.1);
        }
        \node[left] at (-1.5,0.7) {$\xi_R$};
          \foreach \x in {0,...,3} {
            \draw (1.5 , 0.3 + \x) -- (0.5, 0.3 + \x + 1);
            \draw[fill=black] (1.5 ,0.3 + \x) circle (0.1);
            \draw[fill=black] (0.5, 0.3 + \x + 1) circle (0.1);
        }
            \node[right] at (1.5,0.3) {$\sqrt{1-\xi_R^2}$};
    
    \fill[white] (-2,3.5) rectangle (2,4.5);
    \draw (0,0) -- (0,3.5); 
  \node[above] at (-1, 3.5) {$+$};
\node[above] at (1, 3.5) {$-$};
        \draw[red,<-] (-2.5,2.5) --(-2.5,1.5);
\draw[red,->] (2.5,2.5) --(2.5,1.5);
\draw[blue,->] (-3,2.5) --(-3,1.5);
\draw[blue,<-] (3,2.5) --(3,1.5);
    \end{tikzpicture}


V: \quad \scalebox{0.75}{

 \begin{tikzpicture}[baseline={(0,1.5)}]
    \fill[gray!10] (-2,1) rectangle (2,3.5);
    \draw (-2,1) -- (2,1);
    \draw[dashed] (-2,0) -- (2,0);
    
      \foreach \x in {0,...,2} {
            \draw (-1.5 ,  1 + \x) -- (-0.5, 1 + \x + 1);
            \draw[fill=black] (-1.5 ,1 + \x) circle (0.1);
            \draw[fill=black] (-0.5, 1 + \x + 1) circle (0.1);
        }
          \foreach \x in {1,...,3} {
            \draw (1.5 ,  + \x) -- (0.5,  + \x + 1);
            \draw[fill=black] (1.5 , + \x) circle (0.1);
            \draw[fill=black] (0.5,  + \x + 1) circle (0.1);
        }

         \draw[dotted] (0.5,1) -- (1.5,0);
    \draw[fill=gray!5, dotted] (0.5,1) circle (0.1);
    \draw[fill=gray!5, dotted] (1.5,0) circle (0.1);
    
    \fill[white] (-2,3.5) rectangle (2,4.5);
    \draw (0,0) -- (0,3.5); 
\node[above] at (-1, 3.5) {$+$};
\node[above] at (1, 3.5) {$-$};

    \end{tikzpicture}
    \quad 
    \begin{tikzpicture}[baseline={(0,1.5)}]
    \fill[gray!10] (-2,1) rectangle (2,3.5);
    \draw (-2,1) -- (2,1);
    \draw[dashed] (-2,0) -- (2,0);
    
      \foreach \x in {1,...,3} {
            \draw (-1.5 ,  + \x) -- (-0.5,  + \x + 1);
            \draw[fill=black] (-1.5 , + \x) circle (0.1);
            \draw[fill=black] (-0.5,  + \x + 1) circle (0.1);
        }
         \draw[dotted] (-0.5,1) -- (-1.5,0);
    \draw[fill=gray!5, dotted] (-0.5,1) circle (0.1);
    \draw[fill=gray!5, dotted] (-1.5,0) circle (0.1);
        
    \foreach \x in {0,...,2} {
            \draw (1.5 , 1 + \x) -- (0.5, 1 + \x + 1);
            \draw[fill=black] (1.5 ,1 + \x) circle (0.1);
            \draw[fill=black] (0.5, 1 + \x + 1) circle (0.1);
        }
    
    \fill[white] (-2,3.5) rectangle (2,4.5);
    \draw (0,0) -- (0,3.5); 
  \node[above] at (-1, 3.5) {$+$};
\node[above] at (1, 3.5) {$-$};
 
    \end{tikzpicture} 
 \ {\Huge $\leadsto$} \
 \begin{tikzpicture}[baseline={(0,1.5)}]
    \fill[gray!10] (-2,1) rectangle (2,3.5);
    \draw (-2,1) -- (2,1);
    \draw[dashed] (-2,0) -- (2,0);
    
      \foreach \x in {0,...,3} {
            \draw (-1.5 ,   + \x) -- (-0.5,  + \x + 1);
            \draw[fill=black] (-1.5 , + \x) circle (0.1);
            \draw[fill=black] (-0.5,  + \x + 1) circle (0.1);
        }
          \foreach \x in {1,...,2} {
            \draw (1.5 , 1 + \x) -- (0.5, 1 + \x + 1);
            \draw[fill=black] (1.5 , 1 + \x) circle (0.1);
            \draw[fill=black] (0.5, 1  + \x + 1) circle (0.1);
        }

         \draw[dotted] (0.5,2) -- (1.5,1);
    \draw[fill=gray!5, dotted] (0.5,2) circle (0.1);
    \draw[fill=gray!5, dotted] (1.5,1) circle (0.1);
    
    \fill[white] (-2,3.5) rectangle (2,4.5);
    \draw (0,0) -- (0,3.5); 
\node[above] at (-1, 3.5) {$+$};
\node[above] at (1, 3.5) {$-$};

    \end{tikzpicture}
    \quad 
    \begin{tikzpicture}[baseline={(0,1.5)}]
    \fill[gray!10] (-2,1) rectangle (2,3.5);
    \draw (-2,1) -- (2,1);
    \draw[dashed] (-2,0) -- (2,0);
    
      \foreach \x in {2,...,3} {
            \draw (-1.5 ,  + \x) -- (-0.5,  + \x + 1);
            \draw[fill=black] (-1.5 , + \x) circle (0.1);
            \draw[fill=black] (-0.5,  + \x + 1) circle (0.1);
        }
         \draw[dotted] (-0.5,2) -- (-1.5,1);
    \draw[fill=gray!5, dotted] (-0.5,2) circle (0.1);
    \draw[fill=gray!5, dotted] (-1.5,1) circle (0.1);
        
    \foreach \x in {0,...,3} {
            \draw (1.5 ,  + \x) -- (0.5,  + \x + 1);
            \draw[fill=black] (1.5 , + \x) circle (0.1);
            \draw[fill=black] (0.5,  + \x + 1) circle (0.1);
        }
    
    \fill[white] (-2,3.5) rectangle (2,4.5);
    \draw (0,0) -- (0,3.5); 
  \node[above] at (-1, 3.5) {$+$};
\node[above] at (1, 3.5) {$-$};
 
    \end{tikzpicture} 
}


A: \quad \scalebox{0.75}{

 \begin{tikzpicture}[baseline={(0,1.5)}]
    \fill[gray!10] (-2,1) rectangle (2,3.5);
    \draw (-2,1) -- (2,1);
    \draw[dashed] (-2,0) -- (2,0);
    
      \foreach \x in {0,...,2} {
            \draw (-1.5 ,  1 + \x) -- (-0.5, 1 + \x + 1);
            \draw[fill=black] (-1.5 ,1 + \x) circle (0.1);
            \draw[fill=black] (-0.5, 1 + \x + 1) circle (0.1);
        }
          \foreach \x in {1,...,3} {
            \draw (1.5 ,  + \x) -- (0.5,  + \x + 1);
            \draw[fill=black] (1.5 , + \x) circle (0.1);
            \draw[fill=black] (0.5,  + \x + 1) circle (0.1);
        }

         \draw[dotted] (0.5,1) -- (1.5,0);
    \draw[fill=gray!5, dotted] (0.5,1) circle (0.1);
    \draw[fill=gray!5, dotted] (1.5,0) circle (0.1);
    
    \fill[white] (-2,3.5) rectangle (2,4.5);
    \draw (0,0) -- (0,3.5); 
\node[above] at (-1, 3.5) {$+$};
\node[above] at (1, 3.5) {$-$};

    \end{tikzpicture}
    \quad 
    \begin{tikzpicture}[baseline={(0,1.5)}]
    \fill[gray!10] (-2,1) rectangle (2,3.5);
    \draw (-2,1) -- (2,1);
    \draw[dashed] (-2,0) -- (2,0);
    
      \foreach \x in {1,...,3} {
            \draw (-1.5 ,  + \x) -- (-0.5,  + \x + 1);
            \draw[fill=black] (-1.5 , + \x) circle (0.1);
            \draw[fill=black] (-0.5,  + \x + 1) circle (0.1);
        }
         \draw[dotted] (-0.5,1) -- (-1.5,0);
    \draw[fill=gray!5, dotted] (-0.5,1) circle (0.1);
    \draw[fill=gray!5, dotted] (-1.5,0) circle (0.1);
        
    \foreach \x in {0,...,2} {
            \draw (1.5 , 1 + \x) -- (0.5, 1 + \x + 1);
            \draw[fill=black] (1.5 ,1 + \x) circle (0.1);
            \draw[fill=black] (0.5, 1 + \x + 1) circle (0.1);
        }
    
    \fill[white] (-2,3.5) rectangle (2,4.5);
    \draw (0,0) -- (0,3.5); 
  \node[above] at (-1, 3.5) {$+$};
\node[above] at (1, 3.5) {$-$};
 
    \end{tikzpicture} 
 \ {\Huge $\leadsto$} \
 \begin{tikzpicture}[baseline={(0,1.5)}]
    \fill[gray!10] (-2,1) rectangle (2,3.5);
    \draw (-2,1) -- (2,1);
    \draw[dashed] (-2,0) -- (2,0);
    
      \foreach \x in {0,...,3} {
            \draw (-1.5 ,   + \x) -- (-0.5,  + \x + 1);
            \draw[fill=black] (-1.5 , + \x) circle (0.1);
            \draw[fill=black] (-0.5,  + \x + 1) circle (0.1);
        }
          \foreach \x in {1,...,2} {
            \draw (1.5 , 1 + \x) -- (0.5, 1 + \x + 1);
            \draw[fill=black] (1.5 , 1 + \x) circle (0.1);
            \draw[fill=black] (0.5, 1  + \x + 1) circle (0.1);
        }

         \draw[dotted] (0.5,2) -- (1.5,1);
    \draw[fill=gray!5, dotted] (0.5,2) circle (0.1);
    \draw[fill=gray!5, dotted] (1.5,1) circle (0.1);
    
    \fill[white] (-2,3.5) rectangle (2,4.5);
    \draw (0,0) -- (0,3.5); 
\node[above] at (-1, 3.5) {$+$};
\node[above] at (1, 3.5) {$-$};

    \end{tikzpicture}
    \quad 
    \begin{tikzpicture}[baseline={(0,1.5)}]
    \fill[gray!10] (-2,1) rectangle (2,3.5);
    \draw (-2,1) -- (2,1);
    \draw[dashed] (-2,0) -- (2,0);
    
      \foreach \x in {0,...,3} {
            \draw (-1.5 ,  + \x) -- (-0.5,  + \x + 1);
            \draw[fill=black] (-1.5 , + \x) circle (0.1);
            \draw[fill=black] (-0.5,  + \x + 1) circle (0.1);
        }
        
    \foreach \x in {2,...,3} {
            \draw (1.5 ,  + \x) -- (0.5,  + \x + 1);
            \draw[fill=black] (1.5 , + \x) circle (0.1);
            \draw[fill=black] (0.5,  + \x + 1) circle (0.1);
        }
    
    \fill[white] (-2,3.5) rectangle (2,4.5);
    \draw (0,0) -- (0,3.5); 
  \node[above] at (-1, 3.5) {$+$};
\node[above] at (1, 3.5) {$-$};
 
    \end{tikzpicture} 
}

    \end{center}
    \caption{Above, the combined spectral sea for the Dirac fermion. The left and right Weyl fermion copies have independent mixing parameters $\xi_L$ and $\xi_R$. The red and blue arrows indicate the direction of the spectral flow for the vector and axial symmetries, respectively. Below, the spectral sea for the trivial defect and its spectral flow for the vector (V) and axial (A) symmetries.}
    \label{fig: spectraldirac}
\end{figure}
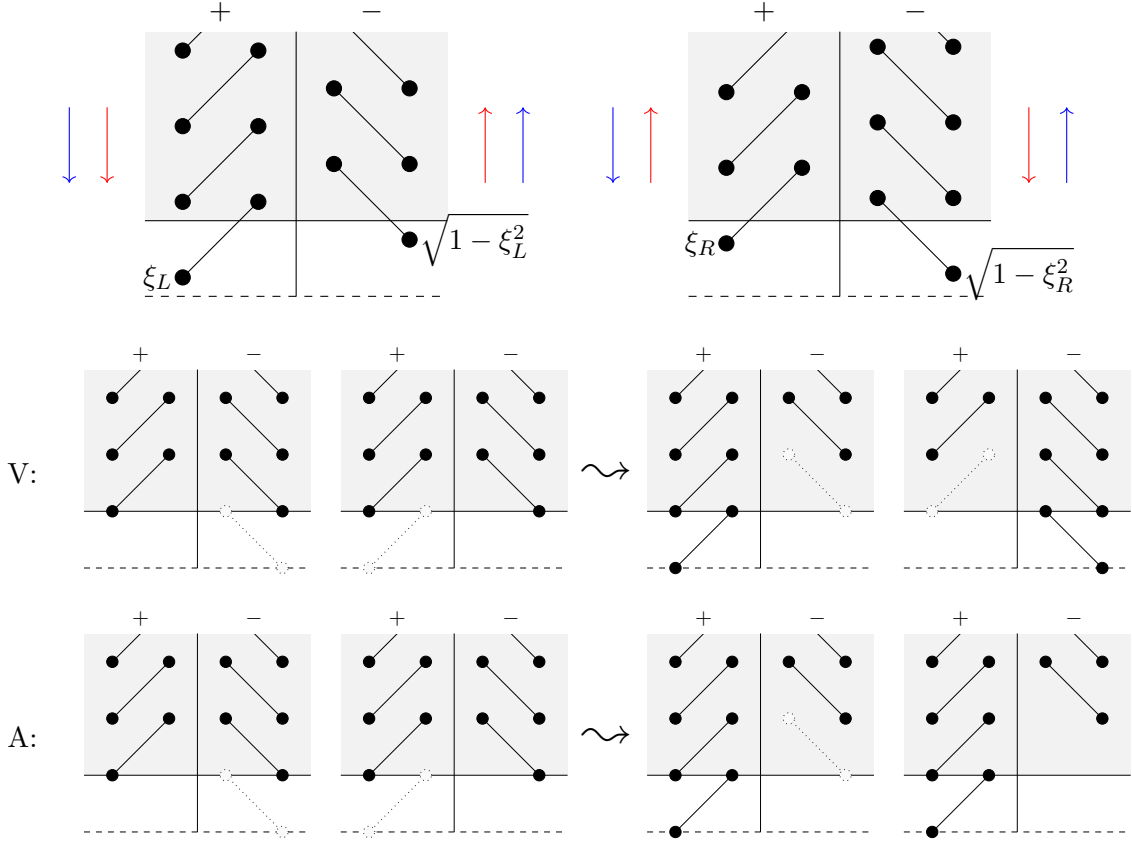

\paragraph{Dirac fermion} In the case of a Dirac fermion, we have two Weyl fermions with independent twists $\nu_L, \nu_R$ and opposite chirality. 
The chirality flips $m \to -m$ and exchanges the two columns of the spectral sea. In Figure \ref{fig: spectraldirac} we show the combined sea for the Dirac fermion.
Let us discuss the vector and axial spectral flow. We start from the trivial defect $\xi_L = 1$ and $\xi_R = 0$. In the case of the vector symmetry, the spectral flow accumulates left- and right-moving zero modes $\chi_L^+, \, \chi_R^-$ on the defect. This leads to a jump $\Delta c_L = \Delta c_R =1$, predicted in \cite{Bianchi:2021snj}. These modes are, however, not resilient. Their conformal dimension is:
\be
\Delta_+ = \Delta_- = 3/2 - \nu \, .
\ee
After $\nu = 1/2$, we can pair up these modes via a relevant mass deformation:
\be
\cO_{\text{mass}} = {\chi_L^+}{}^\dagger \chi_R^- + \text{c.c.} \, ,
\ee
which lifts the zero modes and drives the system back to the trivial defect phase. 

On the other hand, in the case of the axial symmetry, both $L$ and $R$ seas lead to the flow of left-moving modes $\chi_L^+$ and $\chi_R^+$. Thus $\Delta c_L = 2$, $\Delta c_R = 0$. Furthermore, the modes are resilient, as there is no quadratic deformation we can use to lift them. As we have argued in this paper, the modes are resilient due to the bulk 't Hooft anomaly. 

\paragraph{Time reversal and $\nu=1/2$} In the previous section, we have provided two possible realizations of the $\text{Pin}^+_c$ anomaly at $\nu=1/2$.
Let us analyze the case of a single Weyl fermion. Here the time reversal symmetry acts as:
\be
T \chi = (i Y) \chi^* \, ,
\ee
The monodromy defect $\mon{1/2}$ explicitly breaks this symmetry unless $\xi = 1/2$. In the former case, the defect can be made $T$ symmetric at the price of being non-simple e.g.
\be
\mon{\pi} = \mon{\pi}_{\xi = 0} \oplus \mon{\pi}_{\xi=1} \, .
\ee
In the latter case the left and right singular modes are exchanged by $T$ and perfectly degenerate. The same is true for the full $m>0$ and $m<0$ towers. Thus, we conclude that the defect with $\xi = 1/2$ spontaneously breaks the $T$ symmetry, instead of realizing the chiral modes of the T-Pfaffian.
The exact same situation takes place in the case of a Dirac fermion.

\subsection{Spectral flow from the b-function} \label{ssec: bfunc}
The aforementioned structure can be verified explicitly by computing defect central charges.
We will compute the change in the $b$-function of the monodromy defect as we pump a unit of flux $\nu \to \nu + 1$.
This is defined as the coefficient of the (1+1)d Weyl anomaly which resides on the defect worldvolume $\Sigma$:
\be
\delta_\sigma \log(Z_{\calD}) = \frac{b}{24 \pi} \int_{\Sigma} d^2 x \sqrt{h} R_\Sigma \, .
\ee
In \cite{Bianchi:2021snj}, the $b$-function for the vector monodromy defect was computed by integrating the one-point function of the current operator $J_\theta$. The result is:
\be
b_V(\nu, \xi) = \nu^2 (2 -\nu^2 - 2 \xi(3-2 \nu))  +\xi \, ,
\ee
where $\xi_L = \xi_R = \xi$ is a vector-like mixing for the zero modes.
The b-function for a single chiral mode can be extracted from this by recalling that the modes are decoupled, and the symmetry $\nu \to 1 - \nu$, $\xi \to 1 - \xi$ that exchanges chirality and the sign of the twist.
This implies that:
\be
b_L (\nu\, , \xi ) = b_R(1-\nu, 1-\xi) = \frac{1}{2} b_V(\nu, \xi) \, .
\ee
Thus:
\be
b_A(\nu, \xi_L, \xi_R) =  b_L(\nu, \xi_L) + b_R(1-\nu, 1-\xi_R) \, .
\ee
The vector and axial b-functions are multi-valued as $\nu \to \nu + 1$:
\be
b_V(1,0) - b_V(0,0) = 1 \, , \quad b_A(1,0,0) - b_A(0,0,0) = 1 \, .
\ee
As the monodromy defect at $\nu=1$ is also a genuine defect, we can compute the change in its left and right central charges by:
\be
\Delta c_L = \Delta b + \Delta k_g/2 \, , \qquad \Delta c_R = \Delta b - \Delta k_g/2 \, ,
\ee
where $k_g$ is the coefficient of the gravitational anomaly on the defect.
For the vector monodromy defect $\Delta k_g=0$. The pumped degrees of freedom are thus non-chiral and we recover $\Delta c_L = \Delta c_R = 1$. Furthermore, recalling our discussion of defect RG flows, the stable defect is actually given by $\mon{\nu,0}$ for $\nu < 1/2$ and $\mon{\nu,1}$ for $\nu > 1/2$. Piecing together the two b-functions:
\be
b_{\text{stable}}(\nu,\xi) = \begin{cases}
    b(\nu,\xi) \, \quad &0 \leq \nu \leq 1/2 \\
    b(1-\nu, 1-\xi) \, &1/2 < \nu \leq 1 ,
\end{cases}
\ee
we find a continuous and single valued b-function.
Thus, the pumped modes on the vector monodromy defect are not protected.

In the axial case, instead, we have $\Delta k_g=2$ and the change in the central charges is:
\be
\Delta c_L = 2 \, , \qquad \Delta c_R = 0 \, .
\ee
As we have argued earlier, these modes are resilient and anomaly-induced, with chiral central charge $c_- = k_g$ fixed by our arguments.

\subsection{Defect modes, anyons and QED} 
We now discuss the relation between the defect singular modes and the anyons of $\mathbf{T}(\nu)$, and the conditions under which the two are decoupled (or not). 

First, at generic $\nu$, the two are decoupled: anyons are nonlocal objects extending through $\mathbf{T}(\nu)$, and thus cannot appear in the OPE of any genuine bulk fermion. 

Another way to excite the anyonic modes is to study the system in a background monopole configuration. Due to the symmetry fractionalization this excites a fractionally charged anyon on the $\mon{\nu}$ worldvolume. The charge fractionalization can be read-off directly from the fractional Hall response. 
Likewise, we can consider the dimensional reduction on $S^2$ threaded with a unit of magnetic flux. This leads to a (1+1)d fermionic CFT, and the charge fractionalization can be read-off by the mixed vector-axial anomaly there.

A more interesting situation takes place if we instead study QED, i.e. we make the vector symmetry dynamical, denoting its gauge field by $a$.
In this case, it is by now well-known that $U(\nu) \times \mathbf{T}(\nu)$ becomes a non-invertible duality defect $\mathbf{D}(\nu)$ for rational $\nu$ \cite{Choi:2021kmx,Kaidi:2021xfk}.  
The theory $\mathbf{T}(\nu)$ now explicitly interacts with the bulk gauge field by coupling to its field strength. This was the statement of symmetry fractionalization in the ungauged theory.

The gauge theory hosts several dynamical line operators. Wilson lines:
\be
W_n[\gamma] = P\exp\left( i n \int_{\gamma} a \right) = \exp\left( i n \int_{\Sigma: \partial \Sigma = \gamma} d a \right) \, ,
\ee
and magnetic ('t Hooft) lines:
\be
T_k[\gamma] = \exp\left( i k \int_{\Sigma : \partial \Sigma = \gamma} \star d a \right) \, . 
\ee
While the former can terminate on bulk Dirac fields, the latter cannot, as they carry charge under the magnetic one-form symmetry of the theory. 
The situation is radically different in the presence of a duality monodromy defect $\mon{\nu}_{\mathbf{D}}$. Let us consider $\mathbf{D}(1/N)$ for simplicity. This operator acts nontrivially on line defects by 
\be
T_{k} \mathbf{D}(1/N) = \mathbf{D}(1/N) W_{k/N} \, ,
\ee
mapping a 't Hooft line into an ill-quantized Wilson line (i.e. a monodromy defect for the magnetic one-form symmetry). In the presence of a monodromy defect, both objects can terminate on its worldvolume. The fractional gauge charge of the Wilson line is precisely matched by the anyonic excitations on $\mon{1/N}_{\mathbf{D}}$.
Thus, in the QED theory, the anyonic sector is not completely decoupled from the bulk and provides dynamical terminations of magnetic line operators, which only exist on the monodromy defect.

\subsection{A fractional axion string} 
To conclude, we study another classic long distance model which allows to match the chiral anomaly. This is provided by a (3+1)d axion $\phi$ \cite{Antinucci:2024bcm,Antinucci:2025fjp,Seiberg:2025bqy,Seifnashri:2026ema}. The chiral symmetry $U(1)_A$ is transmuted into the shift symmetry $\phi \to \phi + \alpha$ while the anomaly is matched by a background axion coupling:
\be
S = \frac{f^2}{2} \int d^4 x \, \partial_\mu \phi \partial^\mu \phi + \frac{i k}{4 \pi^2} \int \phi \, d A_V \wedge dA_V + i 2 k_g \int \phi \, \widehat{A}(R) \, , \quad \phi \simeq \phi + 2 \pi
\ee
Notice that, even though the IR theory is bosonic, the vortex core remembers the UV Spin$_c$ structure. This can also be seen by the structure of the matched anomaly, which contains a gravitational piece that is ill-defined on its own without the Spin$_c$ structure.
The case we will study here will have $k = k_g = 1$ corresponding to a UV Dirac fermion.

We again consider a monodromy defect $\mon{\nu}$ for the axial symmetry by expanding in cylindrical coordinates \eqref{eq: cylcol} around a background:
\be \label{eq: profile}
d \phi = 2 \nu d \theta \, , \qquad \nu = p/N  \, .
\ee
The factor of $2$ is a convention to match the UV completion of this symmetry via the Jackiw-Rossi model \cite{Jackiw:1981ee}. We will comment on this below.
It is well appreciated that, in the theory with the axion alone, this naive vortex configuration has logarithmically divergent energy. This can be regulated by embedding the axion model into a UV complete theory, with the UV fields confined into the vortex center.

First, let us look at the IR description. Substituting the profile \eqref{eq: profile} in the axion coupling gives rise to an ill-quantized Chern-Simons term:
\be
\frac{i p }{2 \pi N } \int A_V d A_V + i p/N \int \text{CS}_g \, .  
\ee
Once again, we interpret this as stemming from the response of a FQH system. Notice however that, in this case, the FQH modes are completely decoupled from the axion, as perturbations above the winding background are independent of $\nu$.

To elucidate the physics of this mode we must consider a UV-completion of the model:
\be
S_{\text{UV}} = \int d^4x \left( \partial_\mu \Phi \partial^\mu \Phi + i \overline{\Psi} \gamma^\mu \partial_\mu \Psi - V(|\Phi|^2) + (\Phi \overline{\Psi} P_R \Psi + \text{c.c.}) \right) \, , \Psi = (\psi_L, \psi_R) \, .
\ee
$\Phi = f e^{i \phi}$ is a complex scalar, which condenses due to the potential:
\be
\langle |\Phi|^2 \rangle = R^2 \neq 0 \, ,
\ee
giving mass to the fermion and leading to the IR axion.
The UV theory is invariant under an $(U(1)_V \times U(1)_A)/\bZ_2$ symmetry:
\bea
U(1)_V :  &\Psi \to e^{2 \pi i \mu} \Psi \, , \quad &&\Phi \to \Phi \, , \\
U(1)_A : &\Psi \to e^{-2 \pi i \nu \gamma_5 } \Psi \, , \quad &&\Phi \to e^{-4 \pi i \nu}\Phi \, .
\eea
Now we can analyze the vortex configuration in the UV-completed model. The twisted boundary conditions are:
\bea
&\Phi(\theta+2\pi,x,t,r) = e^{-4 \pi i \nu} \Phi(\theta,x,t,r) \, \\
&\Psi(\theta+2\pi,x,t,r) = -e^{-2 \pi i \nu \gamma_5} \Psi(\theta,x,t,r) \, .
\eea
Notice that the fermion b.c. are also twisted. 
The scalar can be treated as a background, which develops a radial profile in the presence of the vortex:
\be
\Phi = f(r) e^{- 2 i \nu \theta} \, ,   \quad f(r) \overset{r \to 0}{\longrightarrow} r^{| 2 \nu |} \, .
\ee
In genuine vortices the field must be single-valued, so $2\nu \in \bZ$ (the standard vortex corresponding to $\nu=1/2$), but this is not the case for the monodromy defect. However notice that, for any nonzero $\nu$, the vortex solution is still regular at the origin and square-integrable. 

Outside the vortex the fermion is massive and can be integrated out. Inside the core we treat $f(r) \sim 0$ and we are left with a massless twisted Dirac fermion, as in Section \ref{sec: chiralmon}. 

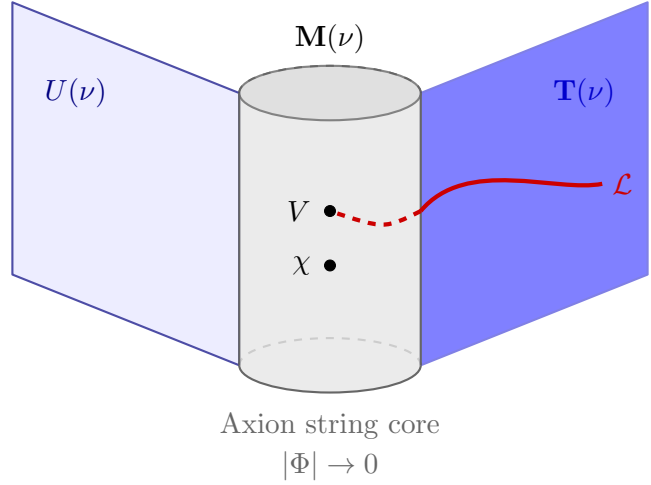
\begin{figure}
    \centering
    \begin{tikzpicture}[scale=1.2]
        \draw[gray!80!black, thick, dashed] (1,-0.5) arc (0:180:1 and 0.3);
        \draw[gray!80!black, thick, dashed] (1,2.5) arc (0:180:1 and 0.3);

        \filldraw[blue!10, opacity=0.7, draw=blue!50!black, thick, line join=round] 
            (-1, -0.5) -- (-3.5, 0.5) -- (-3.5, 3.5) -- (-1, 2.5) -- cycle;
        \node[blue!50!black] at (-2.8, 2.5) {$U(\nu)$};

        \filldraw[blue, opacity=0.5, draw=blue!80!black, thick, line join=round] 
            (1, -0.5) -- (3.5, 0.5) -- (3.5, 3.5) -- (1, 2.5) -- cycle;
        \node[blue!80!black] at (2.8, 2.5) {$\mathbf{T}(\nu)$};

        \fill[gray!20, opacity=0.8] (-1,-0.5) arc (180:360:1 and 0.3) -- (1,2.5) arc (360:180:1 and 0.3) -- cycle;
        \draw[gray!80!black, thick] (-1,-0.5) -- (-1,2.5);
        \draw[gray!80!black, thick] (1,-0.5) -- (1,2.5);
        \draw[gray!80!black, thick] (-1,-0.5) arc (180:360:1 and 0.3);
        \filldraw[gray!30, opacity=0.8, draw=gray!80!black, thick] (0,2.5) ellipse (1 and 0.3);

        \node[black] at (0, 3.1) {$\mon{\nu}$};

        \draw[ultra thick, red!80!black] (3.0, 1.5) .. controls (2.5, 1.4) and (1.5, 1.8) .. (1, 1.2);
        \node[red!80!black, right] at (3.0, 1.5) {$\mathcal{L}$};

        \draw[ultra thick, red!80!black, dashed] (1, 1.2) .. controls (0.6, 1.0) .. (0, 1.2);
        
        \filldraw[black] (0, 1.2) circle (0.06);
        \node[black, left] at (-0.1, 1.2) {$V$};

        \filldraw[black] (0, 0.6) circle (0.06);
        \node[black, left] at (-0.1, 0.6) {$\chi$};

        \node[gray!80!black, align=center] at (0, -1.4) {Axion string core \\ $|\Phi| \to 0$};
    \end{tikzpicture}
    \caption{Chiral modes bound to a fractional axion string: inside the core the UV fermions are deconfined and have chiral zero modes, which attach to the topological lines of the dressing theory $\mathbf{T}(\nu)$. In the IR, these zero modes are decoupled from the axion sector.}
    \label{fig: axionstring}
\end{figure}

Then, we can simply recycle our results and conclude that the vortex generically supports two singular chiral modes,
\be
\chi_L^+ \, , \quad \chi_R^+ \, ,
\ee
localized in its core. We conclude that, in the axion model, the defect modes are really the Jackiw-Rossi fermions confined inside the vortex core, together with anyons attached to the topological lines of the dressing TQFT $\mathbf{T}(\nu)$, see Figure \ref{fig: axionstring}.
Let us contrast the analysis here with the results of \cite{Jackiw:1981ee,Callan:1984sa}. From an anomaly-inflow perspective, the theory on the vortex at $\nu = 1/2$ must soak up the anomaly of a $U(1)_1$ Spin$_c$ SPT. This is correctly reproduced by our computation. However, in contrast with the standard vortex, our fermions are \emph{periodic} around its core due to the $\nu = 1/2$ monodromy. This means that, in order to preserve time reversal, the defect here will be non-simple. On the other hand, in \cite{Jackiw:1981ee} the fermions are anti-periodic and only the scalar twists around the vortex, and lead to a decoupled zero mode matching the $U(1)_1$ Hall response.

\section{Monodromy operators and defects on the lattice} \label{sec: lattice}
To conclude, we briefly discuss the definition of monodromy defects in the context of quantum lattice models.

On the lattice we cannot leverage Lorentz invariance, and so we must explicitly distinguish between the monodromy defect ($\mon{g}$), describing a codimension-2 impurity in space, and a monodromy operator ($\cM(g)$), which acts at a fixed time $t$. The latter can also be interpreted as a defect $\cU(g)$, but for the symmetry generator $U(g)$ instead. An in-depth discussion of the differences between operators and defects in the case of symmetries can be found in \cite{Seifnashri:2023dpa}.

For continuous symmetries the realization of chiral symmetries is notoriously subtle \cite{Nielsen:1980rz,Nielsen:1981hk}, unless we work with an infinite-dimensional on-site Hilbert space.
A well-known example is the (modified) Villain model \cite{Villain:1974ir,Sulejmanpasic:2019ytl,Gorantla:2021svj,Fazza:2022fss}, which thus provides a natural playground for our ideas. 

We now briefly delve into the definition of the monodromy operators and defects on the lattice, and provide examples in the case of anomalous symmetries.

\paragraph{Monodromy operators and the twisted Hilbert space} Consider a symmetry operator $U(g)$ inserted at fixed time. We define a monodromy \emph{operator}, $\cM(g)$ on a codimension-1 spatial surface $\Sigma$ by terminating the topological defect $U(g)$ along $\Sigma$ \cite{Kadanoff:1970kz, Fradkin:1980jn, Dixon:1986qv}, $U_\Sigma(g)$, and setting:
\be
\cM_\Sigma(g) \equiv U_\Sigma(g) \, .
\ee
The truncation is not unique: different choices lead to different monodromy operators. This is not surprising: we already know from the continuum analysis that the monodromy defects are not unique.

The monodromy operators, however, do not act within the lattice Hilbert space $\cH$, but rather provide a map between the original $\cH$ and the twisted Hilbert space $\cH_\Sigma(g)$:
\be 
\cM_\Sigma(g)    :=
\begin{tikzpicture}[baseline={(0,-0.5)}, x={(-0.5cm,-0.4cm)}, y={(1cm,0cm)}, z={(0cm,1cm)}]
    \foreach \x in {0,...,5} {
        \draw[gray!50] (\x,0,0) -- (\x,4,0);
    }
    \foreach \y in {0,...,4} {
        \draw[gray!50] (0,\y,0) -- (5,\y,0);
    }
    \filldraw[cyan, opacity=0.3] (2.5,0,0) -- (5,0,0) -- (5,4,0) -- (2.5,4,0) -- cycle;
    \filldraw[cyan, opacity=0.5] (2.5,0,0) -- (2.5,4,0) -- (2.5,4,2) -- (2.5,0,2) -- cycle;     
    \draw[black, line width=1.5pt] (2.5,0,0) -- (2.5,4,0) node[right] {$\Sigma$};
    \node[black] at (3.75,2,0) {$U_\Sigma(g)$};
    \node[black, right] at (2.5,2,1) {$\cU(g)$};
\end{tikzpicture}
\ee
The Hamiltonian $H_\Sigma(g)$ implementing time evolution in $\cH_\Sigma(g)$ can be defined by conjugating the original Hamiltonian $H$ by the monodromy operator:
\be
H_\Sigma(g) = \cM_\Sigma(g)^\dagger H \cM_\Sigma(g) \, .
\ee
This construction preserves some of the underlying topological properties of $U(g)$:
The Hilbert spaces $\cH_\Sigma(g)$ and $\cH_{\Sigma'}(g)$, for homologous $\Sigma$ and $\Sigma'$, are related by a finite-depth unitary map, called the movement operator $V_{\Sigma \to \Sigma'}$\footnote{This has been proven in the case of (1+1)d systems in \cite{Seifnashri:2023dpa}.} and
\be
H_\Sigma(g) =  V_{\Sigma \to \Sigma'}^\dagger \,  H_{\Sigma'}(g) \,  V_{\Sigma \to \Sigma'} \, .
\ee
In the presence of 't Hooft anomalies the process of terminating $U(g)$ (and thus the definition of $\cM(g)$) is subtle. As pioneered in \cite{Else:2014vma} and recently explored in several works \cite{Shirley:2025yji,Tu:2025bqf,Kapustin:2025nju,Kawagoe:2025ldx,Feng:2025qgg,Feng:2025yge}, the terminated operator must be decorated by a nontrivial unitary in order to reproduce the anomaly cocycle.
In our language, the defect Hilbert space must reproduce the structure of the state space underlying $\mathbf{T}(g)$, for example its anomaly-induced degeneracy. 

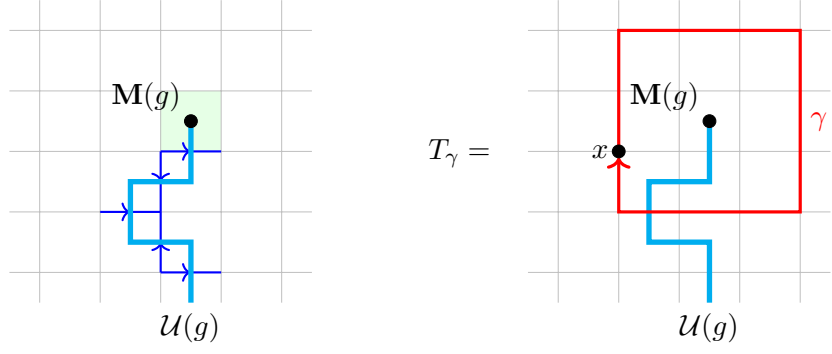
\begin{figure}[t!]
\begin{equation*} 
\begin{tikzpicture}[scale=0.8, baseline={(0,1.5)}]
        \fill[green!10] (2,2) rectangle (3,3);

    \draw[step=1cm, gray!50, very thin] (-0.5,-0.5) grid (4.5,4.5);
    \draw[blue, thick, mid arrow] (2,0) -- (3,0);
    \draw[blue, thick, mid arrow] (2,0) -- (2,1);
    \draw[blue, thick, mid arrow] (1,1) -- (2,1);
    \draw[blue, thick, mid arrow] (2,2) -- (2,1);
    \draw[blue, thick, mid arrow] (2,2) -- (3,2);

    \draw[cyan, line width=2pt] (2.5,-0.5) -- (2.5,0.5) -- (1.5,0.5) -- (1.5,1.5) -- (2.5,1.5) -- (2.5,2.5);
    \filldraw[black] (2.5,2.5) circle (3pt) node[above left] {$\mon{g}$};
    \node[below] at (2.5,-0.5) {$\cU(g)$};
\end{tikzpicture}
\qquad \qquad
T_\gamma = \quad 
\begin{tikzpicture}[scale=0.8, baseline={(0,1.5)}]
    \draw[step=1cm, gray!50, very thin] (-0.5,-0.5) grid (4.5,4.5);
    
    \draw[cyan, line width=2pt] (2.5,-0.5) -- (2.5,0.5) -- (1.5,0.5) -- (1.5,1.5) -- (2.5,1.5) -- (2.5,2.5);
    \filldraw[black] (2.5,2.5) circle (3pt) node[above left] {$\mon{g}$};
    \node[below] at (2.5,-0.5) {$\cU(g)$};
    
    \draw[red, very thick, ->] (1,2) -- (1,4) -- (4,4) -- (4,1) -- (1,1) -- (1,1.9);
    \node[red, right] at (4, 2.5) {$\gamma$};

    \draw[fill=black] (1,2) circle (3pt) node[left] {$x$};
\end{tikzpicture}
\end{equation*}
\caption{Left: The monodromy defect $\mon{g}$ defined as the endpoint of a symmetry defect $\cU(g)$ on the dual lattice, gauge fields $A_\ell = g^{\pm 1}$ are turned on along the marked links. The green plaquette hosts flux for $A_\ell$. Right: The action of the composite translation $T_\gamma$ around a closed loop $\gamma$ on the standard lattice, traversing exactly one dual link.}
\label{fig: monolattice}
\end{figure}

\paragraph{Monodromy defects} 
The monodromy defect $\mon{g}$, instead, can only be defined in more than one spatial dimension, and describes a configuration in which the dynamical impurity extends along time, while being localized at a codimension-2 surface in space. For example, in a (2+1)d system a monodromy defect will be localized e.g. at $x=y=0$ and extend along time. See Figure \ref{fig: monolattice}.

There are two natural ways to describe such a defect. The first is to introduce an external gauge field $A_{\ell} \in \bG$, which assigns a symmetry defect $\cU(A_\ell)$ to a lattice link. Thus, the gauge field must be flat
\be
\prod_{\ell \in P} \exp\left( i A_\ell \right) = 1 \, ,
\ee
where $P$ is a lattice plaquette. On the other hand, at its location, the monodromy defect describes a localized $A_{\ell}$ flux:
\be
\prod_{\ell \in P_{\mon{g}}} \exp\left( i A_\ell \right) = g \, ,
\ee
where $P_{\mon{g}}$ is a plaquette which intersects the location of $\mon{g}$. The monodromy defect is thus identified with a fractional vortex-type configuration on the lattice.
Likewise, we can use the fact that lattice operators $\cO(x)$ in the presence of the monodromy defect $\mon{g}$ satisfy a twisted translation algebra. To see this consider a composite translation $T_\gamma$ which transports a site $x$ around a closed loop $\gamma$ linking $\mon{g}$, as shown in the third panel of Figure \ref{fig: monolattice}.
The translation $T_\gamma$, which would usually be trivial, now acts on operators as the symmetry operator $U(g)$:
\be T_\gamma \, \cO(x) T_\gamma^\dagger = U(g) \, \cO(x) \, U(g)^\dagger \, .\ee
This implies that the translation algebra in the presence of the flux is non-commutative.
To be concrete, if the monodromy defect is localized at a single plaquette $P$, the action of translations on an operator localized at its corner is:
\be
T_x T_y T_x^{-1} T_y^{-1} = U(g) \, .
\ee
We will leave an in-depth study of this description to future work.

\subsection{Monodromy operators and defects in the Villain model}
In this Section we reconsider the effect of a U(1) anomaly on lattice monodromy defects and operators, through the lens of the Villain model. We start in (1+1)d and then move onto the (3+1)d case.
\paragraph{(1+1) dimensional Villain model} The (modified) Villain model is described by the action \cite{Gorantla:2021svj,Seifnashri:2026ema}:
\be
H = \sum_j \frac{1}{2M}  \pi_j^2 + \frac{M}{2} \sum_i \left( \phi_{i+1} - \phi_{i} - w_{i+1,i} \right)^2 \, ,
\ee
The fields $\phi_i \in \bR$ are the Villain scalars, and we have introduced dynamical $\bZ$-valued gauge fields $w_{i+1,i} \in \bZ$ to enforce their compactness. The canonical commutation relations are:
\be
[\pi_i, \phi_j] = i \delta_{ij} \, , \qquad [w_{i+1,i}, b_{j+1,j}] = - i \delta_{ij} \, ,
\ee
where $b_{j+1,j}$ is the canonically conjugated variable to $w_{i+1,i}$, which does not appear in the Hamiltonian.
Furthermore, we impose Gauss-law constraints enforcing the periodicity of the $\phi_i$ fields:
\be
\exp\left( 2 \pi i (\pi_i - b_{i+1,i} + b_{i,i-1}) \right) =1 \, , \qquad \exp\left(2 \pi i w_{j+1,j} \right) = 1 \, .
\ee
This implements the gauge symmetry $\phi_i \to \phi_i + n_i , \, w_{i,i+1} \to w_{i,i+1} + n_{i+1} - n_i$.

The model has a $U(1)_m \times U(1)_w$ symmetry, generated by the charges:
\be
Q_m = \sum_j \pi_j \, , \qquad Q_w =  \sum_j \left( w_{j+1,j} + (\phi_{j+1} - \phi_j) \right)  \, , \label{eq: Villcharge}
\ee
we'll use $q_i$ to denote the quasi-local charges (i.e. $Q = \sum_i q_i$). Notice that we can also globally write $Q_w = \sum_j w_{j+1,j}$ and $Q_m  = \sum_j \left(\pi_j + b_{j+1,j} - b_{j,j-1} \right)$. The two different forms either commute locally with the Hamiltonian, or commute with the local constraints \cite{Seifnashri:2026ema}. We will choose the former in the following discussion.
Let us consider the twisted Hilbert spaces. The truncated symmetry generators are:\footnote{In this Section, we use $U_+$ to denote the truncation to half space.}
\be
U_+(\theta) = \prod_{j=1}^\infty e^{i \theta q_j} \, ,
\ee
and $H(\theta) = U_+(-\theta) H U_+(\theta)$. As commented above, the twisted Hamiltonian remains the same with our choices.
The twisted Hilbert space instead can be interpreted as a modification of the Gauss law constraints on a single link $w_{1,0}$:
\be
e^{2 \pi i w_{1,0}} = e^{i \theta} \, .
\ee
Thus, the $w$ eigenvalue is fractional at the (1,0) link. We can reach this Hilbert space from the untwisted one by acting with a fractionalized vortex operator:
\be
V_\theta = \exp\left( i \frac{\theta}{2\pi} b_{1,0} \right) \, , \qquad  |\psi_\theta\rangle = V_\theta |\psi\rangle \, .
\ee
This is the avatar of the twist field in the continuum model (see Appendix \ref{sec: 1p1}). The winding charge of these states is fractional, and the symmetry realized on the twisted Hilbert space is an extension of $U(1)_w$ by either $\bZ_q$ or $\bR$. Upon performing a $2\pi$ rotation, we see that our states are flowed by the vortex operator $V_{2\pi}$:
\be
|\psi_{2\pi}\rangle = e^{i b_{1,0}} |\psi_0\rangle \, .
\ee
Let us describe the spectral flow in a different manner. Consider the alternative form of the charge:
\be
Q_m = \sum_j \pi_j + b_{j+1,j} - b_{j,j-1}\, .
\ee
With this form, the operator $U_+(\theta)$ keeps the constraints fixed. However the truncated operator now has a nontrivial spectral flow:
\be
U_+(2\pi) = e^{i b_{1,0}} \, U_+(0) \, .
\ee
The operator $e^{i b_{1,0}}$ is a (0+1)d SPT entangler between the trivial and the $U(1)_w$ SPT
\be
\omega(A_w) = e^{i \int A_w} \, .
\ee
This follows from the fact that $e^{i b_{1,0}}$ is charged under $U(1)_w$.

\paragraph{(3+1)d Villain model}
The (3+1)-dimensional version of the Villain model also showcases a similar rich structure \cite{Fidkowski:2025rsq,Thorngren:2026ydw,Lu:2026jnq}. 
We take the Hamiltonian to be:
\be
H = \frac{1}{2M}\sum_{ \bx \, \in \, \text{sites}} \pi_{\bx}^2 + \frac{M}{2} \sum_{\ell \, \in \, \text{links}} \left( d\phi_{\ell} - w_{\ell}\right)^2 + \frac{\lambda}{2} \sum_{P \, \in \, \text{plaquettes}} dw_P^2 \, .
\ee
The last term makes vortices of $w_\ell$ energetically disfavored, while keeping $w_\ell$ non-flat \cite{Lu:2026jnq}.
The canonical commutators are:
\be
[\phi_{\bx}, \pi_{\bx'}] = i \delta_{\bx,\bx'} \, , \qquad [w_\ell, b_{\ell'}] = - i \delta_{\ell, \ell'} \, ,
\ee
and the Hilbert space is subject to Gauss law constraints:
\be
e^{2\pi i \left( \pi_{\bx} - (\nabla \cdot b)_{\bx} \right)} = 1 \, , \qquad e^{2\pi i w_\ell} = 1 \, . 
\ee
The shift symmetry is again on-site and obtained by integrating the momentum operator over space:
\be
Q_m = \sum_{\bx} \pi_{\bx} = \sum_{\bx} \left( \pi_{\bx} - (\nabla \cdot b)_{\bx} \right) \, .
\ee
However, there is also a winding charge
\be
Q_w = \sum_{\Delta_3}  \left( w - d \phi \right) \cup d w = \sum_{\Delta_3} w \cup dw \, .
\ee
Where $\Delta_3$ are fundamental cells of the 3D lattice.\footnote{Alternatively, the model can be defined on any chain complex in 3D, as done e.g. in \cite{Fidkowski:2025rsq,Thorngren:2026ydw}. We will often use their notation as it is more flexible.}

\paragraph{Monodromy operators}
First we study the monodromy operators $\cM(\theta)$. 
The best way to do so is to turn on a gauge field $A_\ell$ for the shift symmetry $U(1)^m$, as shown in \cite{Seifnashri:2026ema}.
The gauge-covariantized charge reads:
\be
Q_w[A] = \int_{\Delta_3} (w - d \phi + \frac{A}{2\pi}) \cup (dw + \frac{dA}{2\pi}) \, .
\ee
As noticed in \cite{Seifnashri:2026ema}, this charge is gauge invariant, but it is not quantized. The lack of quantization can be understood as the anomaly-induced spectral pump, see below.\footnote{If the charge were quantized, then the pump must be trivial as the SPT $\omega(\gamma)$ partition function is a nontrivial phase.}
The monodromy operator is built from the truncated operator:
\be
\begin{aligned}
U_+^w(\theta) = \exp\left[\, i \theta \int_{\Delta_3} \left( w \cup dw + \tfrac{1}{(2\pi)^2} A \cup dA + \tfrac{1}{2\pi} (w \cup dA + A \cup dw) \right) \right. \\
\left. - i \frac{\theta}{2\pi} \int_{\partial \Delta_3} \left( \phi \cup dw - d\phi \cup A \right) \,\right]
\end{aligned}
\ee
This construction is gauge-invariant, however the background term itself is ill-defined, and must be understood as the Hall response of a FQH state. This matches with the continuum analysis.

The spectral pump is relatively easy to see in this description, and does not need a specific Hamiltonian.
Consider the defect Hilbert space $\cH_{2\pi}$, obtained by conjugating the original Hamiltonian by the truncated operator $U_+^w(\theta)$ at $\theta = 2\pi$. Clearly, far from the cut boundary, $U_+^w(2\pi)$ is the trivial unitary, however, taking into account the boundary contribution, we find\footnote{We are using that $w$ is an integer-valued one-form on the lattice.}:
\be
U_+^w(2\pi) = \exp\left(2 \pi i \sum_{\partial \Delta_3^+} \phi \cup d w \right) U_+^w(0)  \, .
\ee
This is a nontrivial operator, which can be recognized as the SPT-entangler for the level-2 $U(1)$ SPT (this can be seen, for example, by turning on a gauge field $A$ explicitly):
\be
\omega(A) = \frac{1}{2\pi} \int A \cup d A \;=\; \frac{2}{4\pi} \int A \, dA \, ,
\ee
reproducing the spectral pump for the fractional axion string.

\paragraph{A $\pi$-flux monodromy defect} Finally, let us provide an example of nontrivial monodromy defect which we can analyze in the Villain model: the $\pi$ flux for the shift symmetry.
Following our discussion, we introduce a $\bZ_2^m \subset U(1)^m$ monodromy defect on the lattice. The symmetry defect $\cU(\pi)$ extends along the $y$ and $z$ directions at $x=0$, and the monodromy defect $\mon{\pi}$ sits at $x=y=0$ and extends only along $z$.

The defect Hamiltonian reads:
\be
H = \frac{1}{2M}\sum_{\bx} \pi_{\bx}^2 + \frac{M}{2} \sum_\ell \left( d \phi_\ell - w_\ell - \frac{A_\ell}{2\pi}  \right)^2 + \frac{\lambda}{2}\sum_{P}\left(d w_P - \frac{d A_P}{2\pi}\right)^2 \, ,
\ee
where the non-vanishing gauge field components can be taken to be $A_{[(0,y,z), (1,y,z)]} = \pi$ for $y\geq 0$. This set of lattice links forms a 2D surface $\Sigma$ inside the 3D lattice where $\cU^m(\pi)$ is placed. The defect's spectrum is doubly degenerate:
\be
\mon{\pi} = \mon{\pi}^+ \oplus \mon{\pi}^- \, ,
\ee 
as both $(dw)_P = \pm 1$ have the same energy, and preserves $[U(1)^m \times U(1)^w]\rtimes C$, where $C$ is charge conjugation. Let us study how these symmetries act on the defect:
\begin{itemize}
    \item The momentum and winding symmetries $U(1)^m \times U(1)^w$ still commute exactly with the defect Hamiltonian, even in the presence of the gauge field.
    \item The charge conjugation symmetry $C$, instead, only commutes up to a shift of $w$ by one unit:\footnote{A similar treatment of (2+1)d vortices is given in \cite{Komargodski:2025jbu}.}
    \be
    \widetilde{C} = C \prod_{\ell \in \Sigma} e^{i b_\ell} \, .  
    \ee
    Since $b_\ell$ is $C$-odd, this operator is still of order 2.
\end{itemize}
Let us analyze the consequences of this structure. Consider the conjugation relation:
\be
\widetilde{C} Q_w \widetilde{C}^\dagger = - Q_w - \int_{\Sigma} d w \, .
\ee
Exponentiating this relation gives:
\be
\widetilde{C} \, U_+^w(\theta) \, \widetilde{C}^\dagger = U_+^w(-\theta)  \, \exp\left( -i \theta \int_{ \Sigma} dw \right) \, .
\ee
As the last term is a total derivative, the naive $O(2)$ structure is modified on the monodromy defect. The effect of this term can be clarified by introducing a $U(1)^m$ gauge field $A_\ell$ on $\mon{\pi}$, which shifts $w$ to $w + A/2\pi$, leading to:
\be
\widetilde{C} \, U_+^w(\theta) \, \widetilde{C}^\dagger \, U_+^w(\theta) = \exp\left( - i \theta  \int_{\partial \Sigma} w - \frac{i \theta}{2 \pi} \int_\Sigma d A \right)
\ee
We claim that this is exactly the structure predicted by the continuum anomaly between $U(1)^m, \, U(1)^w$ and $C$. In the continuum we start from the 6-form anomaly polynomial:
\be
P_6 = \frac{1}{(2 \pi)^2} F_m \wedge F_w \wedge F_w \, .
\ee
On the $\pi$ flux we find the response:
\be
\tau(\pi) = \frac{1}{4\pi} dA_w \wedge dA_m \, ,
\ee
which again describes a nontrivial SPT for $ \left( U(1) \times U(1) \right) \rtimes C$. Using the boundary Hall response and acting with $C$, this implies the relation:
\be
C \, U^w(\theta) = U^w(-\theta) \, C \, \exp\left(- i \frac{\theta}{2\pi} \int_{\Sigma} d A_m \right) \, ,
\ee
which exactly matches the lattice result upon identifying the background gauge field $A_m^{\text{continuum}} = A^{\text{lattice}}$.

\section{Discussion} 
In this work we have examined the imprint of bulk 't Hooft anomaly on dynamical monodromy defects through the lens of decorated domain walls ---which generalize string order parameters--- in massive SPT phases.
This is by no means a complete study, and there are several interesting open directions to pursue: 
\begin{enumerate}
    \item \emph{Chiral modes on the lattice} It would be of great interest to realize the setup of Section \ref{sec: chiralmon} in a fermionic lattice theory. A natural setup is provided by Weyl semimetals \cite{Armitage:2017cjs}, where external axial gauge fields can be introduced via modulations in the UV fermionic lattice \cite{Cortijo:2015hlt}. Similar effects have been predicted as a consequence of dislocations in the Weyl semimetal's lattice \cite{sumiyoshi2016torsional}, see also \cite{ran2009one}.
    \item \emph{Renyi and conical defects} Another class of defects which are closely related to monodromy defects are conical defects, which describe conical singularities in spacetime. This dictionary is precise when the deficit angle is $2\pi$ times an integer. In this case the defect can be reinterpreted in the replica language as a monodromy defect arising from the $\bZ_n$ replica symmetry \cite{Hung:2014npa,Bueno:2015qya,Bueno:2015rda,Bianchi:2015liz,Faulkner:2015csl}. 
    Crucially, a conical geometry sources a nontrivial spacetime holonomy, hinting that gravitational anomalies could have interesting implications for entanglement entropy computations.
    \item  \emph{Higher monodromy pumps:} if $\pi_1(\bG) \neq 0$, we have pointed out that an anomaly in the space of defect couplings leads to a pump of gapless modes on $\mon{g}$. It would be interesting to understand how higher homotopy groups can be detected by $\mon{g}$. 
    \item \emph{SymTFT} Our results can be reinterpreted through the lens of the Symmetry TFT \cite{Bhardwaj:2023ayw,Bartsch:2023pzl,Copetti:2024onh}, which would allow us to extend our observations to monodromy defects for non-invertible symmetries, which have seen little study in dynamical setups.
    \item \emph{Callan-Rubakov and Categorical Scattering} Scattering amplitudes in the presence of magnetic monopoles \cite{Rubakov:1981rg,rubakov1982adler,Callan:1982ah,Callan:1982au,vanBeest:2023dbu,vanBeest:2023mbs,Loladze:2025jsq} suggest that operators in the twisted sectors of (anomalous) bulk symmetries could be excited in the out-going radiation.\footnote{See also \cite{Ueda:2025ecm,Antinucci:2026uuh} for studies of such processes for generic interfaces.} Monodromy defects act in some sense as ``fractional'' monopoles, binding chiral fermionic modes to their core. It is tempting to conjecture that bulk scattering experiments could detect these anomaly-induced modes.
\end{enumerate}

\paragraph{Acknowledgment} The author is grateful to Andrea Antinucci, Gabriel Cuomo, Lorenzo Di Pietro, Giovanni Galati, Chris Herzog, Brandon Rayhaun,  Giovanni Rizi, Shu-Heng Shao and Yifan Wang for discussions. The author is supported by the STFC grant ST/X000761/1. The author would like to thank the Isaac Newton Institute for Mathematical Sciences, Cambridge, for support and hospitality during the program ``Quantum field theory with boundaries, impurities, and defects", the Bernoulli Center for Fundamental Studies for support and hospitality during the workshop ``QFT in AdS 2026", where work on this paper was undertaken; as well as the organizers of the conference ``Symmetry and Interfaces'' for support and hospitality during the event, which has led to several informative discussions.

\appendix

\section{Monodromy defects for the free boson}\label{sec: 1p1} 
We showcase our construction in a (1+1)d setup, where the monodromy defect is described by a local disorder-type operator. This type of symmetry constraint has been discussed prominently in \cite{Chang:2018iay}.
Higher-dimensional examples are treated in Section \ref{sec: chiralmon} and Appendix \ref{sec: 2Group}.

A canonical playground is the free compact boson $X \sim X + 2 \pi$, with action:
\be
S = \frac{R^2}{4 \pi} \int  d X \wedge \star d X \, .
\ee

\subsection{The $U(1)^m \times U(1)^w$ symmetry}\label{ssec: u1u1}
The model has a $U(1)^m \times U(1)^w$ symmetry, with a mixed anomaly:
\be
I_4(A_m, A_w) = \frac{1}{2\pi} d A_m \wedge d A_w \, .
\ee
Charged primaries are the vertex operators $V_{n,m}$ of dimension $\Delta = \frac{n^2}{2 R^2} + \frac{m ^2 R^2}{2}$ and spin $s = m n$.

Let us consider a monodromy defect $\mon{\theta_w}$ for the winding symmetry. Our analysis gives us the transgression form:
\be
\tau(\theta_w) = \frac{\theta_w}{2\pi} d A_m \, , \qquad \mathbf{T}(\theta_w) = \exp\left(\frac{i \theta_w}{2 \pi} \int A_m \right) \, .
\ee
Our expression for $\mathbf{T}(\theta_w)$ describes a 1d Chern-Simons term with fractionalized level. For a $U(1)$ symmetry this is ill-defined. In this case, we must interpret it as the statement that the symmetry on the monodromy defect is fractionalized and the correct group to consider is an extension $\bG_{\theta_w}$ of $U(1)$. There are two cases: for irrational $\theta_w$ the symmetry realized on the monodromy defect is $\bR$, corresponding to the extension:
\be
1 \longrightarrow \bZ \longrightarrow \bR \longrightarrow U(1) \longrightarrow 1
\ee
while for rational $\theta_w/2\pi = p/q$, $\gcd(p,q)=1$ we have:
\be
1 \longrightarrow \bZ_q \longrightarrow U(1) \longrightarrow U(1) \longrightarrow 1 \, .
\ee
Interpreting $A_m$ as a $\bG_{\theta_w}$ gauge field, the level of the Chern-Simons term is well-quantized and $\mathbf{T}(\theta_w)$ becomes well-defined. 
As $\pi_1(U(1))=\bZ$ we may also have a nontrivial defect pump. This is indeed what happens and it is simple to check that:
\be
\mathbf{T}(2\pi) = e^{i \int A_m} \mathbf{T}(0) \, .
\ee
This leads to (minus) one unit of momentum charge being pumped on the flux defect after a full $2\pi$ rotation.\footnote{The $(-1)$ factor follows from carefully tracking the orientation of $\mathbf{T}$.} Thus, we expect level crossing on the twisted Hilbert space on the loop.

Let us verify these predictions. The twisted sector operators are ill-quantized vertex operators:
\be
V_{\frac{\theta_w}{2\pi} + \ell, m} \, .
\ee
For $\theta_w \neq 2 \pi k$ the ground state is non-degenerate and every state carries fractional momentum charge $q = \frac{\theta_w}{2\pi} \mod 1$ and the symmetry group is extended in accordance with the prediction above. A $2\pi$ rotation of $\theta_w$ induces a nontrivial spectral flow on the defect Hilbert space. The ground state at $\theta_w = 0$ is described by $V_{0,0}=\unit$, which interpolates at $\theta_w = 2\pi$ to $V_{-1,0}$, which carries charge $-1$ under the momentum symmetry. We conclude that a charge of $-1$ is pumped into the system as $\theta_w$ winds around the circle. 

Yet another derivation of this fact hinges on the conservation equation for the momentum symmetry:
\be
\partial_\mu J^\mu_m = \frac{1}{2\pi} d A_w \, .
\ee
In a monodromy defect background we have $d A_w = \theta_w \, \delta^2(0)$. The 't Hooft anomaly in this background reduces to:
\be
\partial_\mu J^\mu_m = \frac{\theta_w}{2\pi} \, \delta^2(0) \, ,
\ee
which is exactly the equation describing a localized momentum charge $q = \frac{\theta_w}{2\pi}$ at the origin of spacetime.

For generic $\theta_w$, the twisted Hilbert space has a single vacuum. However, for $\theta_w = \pi$, we have an emergent charge conjugation symmetry which is spontaneously broken in $\cH_\pi$, leading to degenerate levels built on top of $V_{+ 1/2, 0}$ and $V_{-1/2,0}$. The Hilbert space thus is a direct sum:
\be
\cH_\pi = \cH^+ \oplus \cH^- \, ,
\ee
as expected from the spectral pump.

\subsection{Type III anomalies} Another example is a (1+1)d system with a $\bZ_2^3$ global symmetry, with a type III anomaly:
\be
\Omega^3(A_1,A_2,A_3) = \pi \int A_1 \cup A_2 \cup A_3 \, .
\ee
This anomaly is realized again in the free compact boson with $\bZ_2^3 = \bZ_2^C  \times \bZ_2^m \times \bZ_2^w$.
Let us consider the twist defect for momentum symmetry $\bZ_2^m$. This corresponds to a non-flat background $A_1$ field. Applying the reduction map gives:
\be
\Omega^m(A_2,A_3)= \pi \int A_C \cup A_w \, ,
\ee
which is a nontrivial element of $H^2(\bZ_2^2, U(1))$ (a nontrivial SPT). Its boundary is described by a topological quantum mechanics with a single qubit Hilbert space $\cH_{\mathbf{T}(\eta_m)} = \bC^2$. On this space, the other two $\bZ_2$ symmetries act as Pauli $X$ and $i Z$ matrices respectively. We conclude that operators in $\cH_{\mathbf{T}(\eta_m)}$ transform projectively under $\bZ_2^C \times \bZ_2^w$. This is easy to verify in the example of the free boson: twisted operators are vertex operators of the form:
\be
V_{n,m} \, ,  n \in \bZ, \, m \in \frac{1}{2} + \bZ \, ,
\ee
The operators $V_{n,m}$ and $V_{-n,-m}$ form a projective representation of $\bZ_2^C \times \bZ_2^w$ on which the symmetries act as $C = X$, $\eta_w = i Z$. The factor of $i$ stems from the type II anomaly between $\bZ_2^w$ and $\bZ_2^m$, $\Omega^{II} = \pi \int A_m \beta(A_w)$. This is the $\bZ_2$ restriction of the discussion in Section \ref{ssec: u1u1} and implies that the winding symmetry is extended to $\bZ_4$ in the $\bZ_2^m$-twisted sector.

Together, Appendix \ref{ssec: u1u1} and the present subsection realize three of the four scenarios of Section \ref{sec: gapless} (spontaneous breaking, group extension, and spectral pump) in a controlled (1+1)d setting. We now turn to a (2+1)d realization of the higher-group-extension scenario.

\section{Fractional Wilson lines and the 2-Group}\label{sec: 2Group}

Let us come to a second example, which concerns the physics of $U(1)_T$ monodromy defects for (2+1)d scalar QED with $N_f > 1$ scalars, with Euclidean action
\be
S = \int d^3 x D_a \Phi D_a \Phi^\dagger + \frac{1}{4 e^2} f^2 + V(|\Phi|) \, .
\ee
This theory has a PSU($N_f$)$\times U(1)_T$ symmetry, where the second factor is generated by the topological current:
\be
\star J_T = \frac{1}{2 \pi} f \, .
\ee
This symmetry gives charge to the magnetic monopoles of the theory.
It is well known \cite{Komargodski:2017dmc} that the two symmetries have a nontrivial mixed 't Hooft anomaly:
\be
\Omega = \frac{2 \pi}{N_f} \int \frac{F}{2 \pi} \cup w_2(\text{PSU}(N_f)) \, ,
\ee
where $w_2$ is the second Stiefel-Whitney class, and $F = dA$ is the background field strength for $U(1)_T$.

We now consider a monodromy defect $\mon{\theta}$ for $U(1)_T$. This provides twisted boundary conditions for monopoles and describes Wilson lines of fractional charge:
\be
\exp\left( i \theta \int f \right) \simeq W_\theta \, .
\ee
Let us restrict our attention to a $\bZ_{N_f} \subset U(1)_T$ twist. The anomaly restricted to the subgroup takes the form:
\be
\Omega = \frac{2 \pi}{N_f} \int A \cup \beta(w_2) \, .
\ee
Our arguments show that:
\be
\tau(k) = \frac{2 \pi k}{N_f} \int \beta(w_2) \, .
\ee
The situation is analogous to the scalar example we have studied before. The term in the integrand is a trivial SPT, as $H^3(B\text{PSU}(N_f),U(1)) = 0$. However, we cannot trivialize it with a PSU($N_f$) symmetric term. 

We again turn to a group extension, now by a 2-Group $\Gamma_{N_f}$:
\be
1 \longrightarrow \bZ_{N_f}^{(1)} \longrightarrow \Gamma_{N_f} \longrightarrow PSU(N_f) \longrightarrow 1 \, .
\ee
Its defining equation is \cite{Benini:2018reh}:
\be
d B =\beta(w_2) \, ,
\ee
where $B$ is a 2-form $\bZ_{N_f}$ gauge field. Thus
\be
\mathbf{T}(k) = \frac{2 \pi k}{N_f} \int B \, ,
\ee
and the fractional Wilson line carries charge $k$ under the emergent 2-group symmetry. This is a higher-form analogue of the compact boson example we have described earlier.

This can be verified by a discrete gauging. Gauging a $\bZ_{N_f}$ subgroup of $U(1)_T$ makes the fractionalized Wilson line into a genuine object. However, due to the mixed anomaly \cite{Tachikawa:2017gyf}, the symmetry of the gauged theory is a 2-group $\Gamma_{N_f}$, under which $W_{k/N_f}$ carries charge $k$.

The monodromy defect also has a nontrivial defect pump under $U(1)$: a $2\pi n$ rotation pumps an SPT:
\be
\omega(n) = \frac{2 \pi n}{N_f} \int w_2 \, .
\ee
Thus, pumping $n$ units of magnetic $U(1)_T$ flux, leads to a fractionalized line operator, whose N-ality under the cover SU($N_f$) is $n$. This is again simple to verify: a $2\pi n$ vortex is a genuine Wilson line $W_n$, which is known to carry a defect anomaly \cite{Delmastro:2022pfo,Brennan:2022tyl,Antinucci:2024izg} $\omega(n)$, leading to a fractionalized transformation under the flavor symmetry.

\section{Normalization of $\theta$ angles for bosonic, Spin, and Spin$_c$ structures} \label{app: thetaperiod}

In several places in the main text we have used the (3+1)d $U(1)$ topological term
\be \label{eq: thetaaction}
\tau(\theta) \;=\; \frac{\theta}{8\pi^2} \int_{M_4} F \wedge F \, , \qquad F = dA \, ,
\ee
treating $\theta$ as a periodic parameter. The period depends crucially on the spacetime structure available on $M_4$. We collect here the three relevant cases and the dictionary between them. The discussion is standard; see e.g.\ \cite{Seiberg:2016rsg} for a thorough treatment.

Throughout, we normalize the background connection $A$ so the first Chern class $c_1 = F/2\pi \in H^2(M_4,\bZ)$ has integer periods, and
\be
\frac{1}{8\pi^2}\int_{M_4} F \wedge F \;=\; \frac{1}{2}\int_{M_4} c_1 \cup c_1 \;\equiv\; \frac{1}{2} N(M_4) \, , \qquad N(M_4) \in \bZ \, .
\ee
The action $e^{i\tau(\theta)} = e^{i\theta\, N(M_4)/2}$ is invariant under $\theta \to \theta+2\pi$ if and only if $N(M_4) \in 2\bZ$ for every admissible $(M_4, A)$.

\paragraph{Bosonic case.} On a generic oriented 4-manifold $c_1$ is unconstrained and $N(M_4)$ can be any integer. Indeed, on $M_4 = \bC P^2$ with $c_1$ the hyperplane class, $N(\bC P^2) = 1$. Hence
\be  \quad \theta \;\sim\; \theta + 4\pi \quad \text{(bosonic)} \quad 
\ee
The non-trivial $\theta = 2\pi$ point is detected on $\bC P^2$ and corresponds to the bosonic Hall SPT.

\paragraph{Spin case.} On a Spin manifold $w_2(M_4) = 0$. The Wu formula in 4d implies $c_1\cup c_1 \equiv c_1\cup w_2 \pmod 2$, hence $N(M_4) \in 2\bZ$. The action \eqref{eq: thetaaction} is therefore well-defined modulo $2\pi$:
\be  \quad \theta \;\sim\; \theta + 2\pi \quad \text{(Spin)} \quad 
\ee
This is the periodicity used in Section~\ref{sec: topomon}.

\paragraph{Spin$_c$ case.} A Spin$_c$ structure is a $U(1)$ line bundle:
\be
c_1(\cL) \equiv w_2(M_4) \pmod 2
\ee 
On a generic Spin$_c$ 4-manifold $\int c_1^2$ can be \emph{odd}: e.g.\ $\bC P^2$ with the standard Spin$_c$ structure has $\int c_1^2 = 1$. Naively this would force $\theta \sim \theta + 4\pi$ as in the bosonic case. The expected periodicity $\theta \sim \theta + 2\pi$ is recovered only after combining the bare action \eqref{eq: thetaaction} with the $\eta$-invariant of the Spin$_c$ Dirac operator (equivalently, with the appropriate gravitational counterterm). Concretely, by Atiyah--Singer
\be \label{eq: spincindex}
\mathrm{Index}\,\slashed{D}_{\mathrm{Spin}_c}(M_4) \;=\; \int_{M_4} \widehat{A}(R)\, e^{c_1} \;=\; \int_{M_4} \left[ \frac{c_1^2}{2} - \frac{p_1(R)}{24}\right]\, \in\, \bZ\, ,
\ee
with $p_1$ the first Pontryagin class of the tangent bundle. Combining \eqref{eq: thetaaction} with the gravitational completion to reproduce $\pi\,\mathrm{Index}$, one finds the well-defined Spin$_c$ topological term
\be \label{eq: spinctheta}
\tau_{\mathrm{Spin}_c}(\theta) \;=\; \theta \left(\frac{1}{2} \int_{M_4} \frac{F}{2\pi} \wedge \frac{F}{2 \pi} - \frac{\sigma}{8} \right)
\ee
where $\sigma = - \frac{1}{24 \pi^2} \int_{M_4} \Tr R \wedge R $ is the signature.
This is invariant under $\theta\to\theta+2\pi$ on any closed Spin$_c$ manifold:
\be  \quad \theta \;\sim\; \theta + 2\pi \quad \text{(Spin}_c\text{)} \quad 
\ee
The $(3{+}1)$d topological insulator sits at the non-trivial $\bZ_2$ point $\theta = \pi$. Notice that, on a spin manifold, the signature is a multiple of $16$ and we can safely absorb the gravitational correction in the gravitational $\theta$ angle.

\section{Minimal Spin$_c$ theories} \label{app: bostospin}
In this appendix we review the $\cA_{N,p}$ theories of \cite{Hsin:2018vcg} and their compatibility with a Spin$_c$ structure. 

The $\cA_{N,p}$ theory is generated by a line $L$ with spin:
\be
\theta_{L^a} = \exp\left( \frac{\pi i p}{N} a^2 \right) \, ,
\ee
and braiding:
\be
S_{ab} = \exp\left(2 \pi i \frac{p}{N} a b \right) \, .
\ee
From this it follows that $L^N$ has spin $(-)^{Np}$ and is a transparent boson for $Np$ even and a transparent fermion $\psi$ for $Np$ odd. Furthermore, $L$ has one-form symmetry charge $p$.
In the latter case the theory needs a Spin structure to be defined, and $p \simeq p + N$. In the former case, the theory is bosonic, and $p\simeq p + 2N$. 

When $N$ is odd, and $p$ is even, the theory can be given a Spin$_c$ structure, as the boson $L^N$ always carries even electric charge for any choice of fractionalization. 
When instead $p$ is odd, and $N$ is even, only even choices of fractionalization are compatible with the Spin$_c$ structure. 

\paragraph{Chiral central charge}
For the bosonic case, the chiral central charge mod 8 is given by the Gauss sum:
\be
e^{2 \pi i c_-/8} = \frac{1}{\sqrt{N}} \sum_{j=0}^{N-1} e^{\frac{2 \pi i p}{2 N} j^2} \, ,
\ee
which can be evaluated to be \cite{Hsin:2018vcg}:
\be
e^{2 \pi i c_-/8} \equiv G_{N,p} = \begin{cases}
    \left( \frac{p/2}{N} \right) \epsilon(N) \, , \quad &N \, \text{odd} \, , p \, \text{even} \, , \\
    \left(\frac{N/2}{p}\right) \epsilon(p)^{-1} e^{i \pi /4} \, , \quad &N \, \text{even} \, , p \, \text{odd} \, . 
\end{cases}
\ee
With $\left(\frac{p}{q} \right)$ the Jacobi symbol and $\epsilon(s)=1$ for $s=1 \mod 4$ and $\epsilon(s)=-1$ for $s= 3 \mod 4$.

For Spin theories the central charge can only be detected mod 1/2. However, for their Spin$_c$ lift, this is not enough, as the invertible Spin TQFTs Spin(n)$_1$ are not compatible with the Spin$_c$ structure.
The correct observable in this case is the $\mathbf{CP}^2$ partition function with $U(1)$ flux, which is given in our case by \cite{Kobayashi:2021jsc}:
\be
G_{N,p;\kappa}(\xi) = \frac{1}{\sqrt{N}}\sum_{[a]} \exp\left( \frac{i \pi p}{N}(a^2 - \kappa a \xi) \right) \, .
\ee
The second  term being the charge of the fractionalized anyon times the (odd) flux $\xi$ and $[a]$ means equivalence modulo fusion with the transparent fermion $\psi$. In particular, for $\xi =1$ this sum leads to the invariant $(c_- - \sigma_H) \mod 8$, and can be used to compute the chiral central charge of the Spin$_c$ minimal theory.

\paragraph{Hall response}
The Hall conductivity can be computed from the anomaly:
\be
\Omega_{N,p} = \frac{2 \pi p}{2 N} \fP(B) \, ,
\ee
which can be re-expressed using continuum fields $B_c$ as:
\be
\Omega_{N,p} = \frac{N p}{4 \pi} B_c \wedge B_c + \frac{N}{2 \pi} B_c \wedge \lambda \, ,
\ee
where $\lambda$ is a dynamical Lagrange multiplier restricting $B_c$ to a $\bZ_N$ gauge field. Plugging in the symmetry fractionalization:
\be
B_c = \frac{\kappa}{N} F \, ,
\ee
one finds the response:
\be
\tau(N,p;\kappa) = \frac{p \kappa^2}{4 \pi N} A \wedge d A \, .
\ee

\small{
\bibliographystyle{ytphys}
\baselineskip=0.85\baselineskip
\bibliography{mybib}
}

\end{document}